\documentclass[twocolumn]{aastex63} 

\usepackage{graphicx} 
\usepackage{float} 
\usepackage{wrapfig} 
\usepackage{lipsum} 
\usepackage{hyperref}
\usepackage{times}
\usepackage{amsmath}
\usepackage{graphicx}
\usepackage{subfigure}
\usepackage{placeins}
\usepackage{hyperref}
\usepackage{gensymb}
\usepackage{upgreek}
\usepackage{natbib}

\usepackage{graphicx}
\usepackage{subfigure}
\usepackage{multirow}
\usepackage{comment}
\usepackage{natbib}
\usepackage{hyperref}
\usepackage{mathtools}
\usepackage{mathrsfs}
\usepackage{fontenc}
\usepackage{color}
\usepackage{url}
\usepackage{hyperref}
\usepackage{gensymb}
\usepackage{pifont}
\graphicspath{{./}}

\newcommand       \K            {\,{\rm K}}
\newcommand       \mum          {{\rm \mu m}}

\usepackage{afterpage}

\shorttitle{PAH Characteristics of GATOS Seyferts}
\shortauthors{Zhang et al.}

\begin{document}

\title{Polycyclic Aromatic Hydrocarbon Emission in the Central Regions of Three Seyferts and the Implication for Underlying Feedback Mechanisms}

\author[0000-0003-4937-9077]{Lulu Zhang}
\affiliation{The University of Texas at San Antonio, One UTSA Circle, San Antonio, TX 78249, USA; lulu.zhang@utsa.edu; l.l.zhangastro@gmail.com}

\author{Ismael Garc{\'i}a-Bernete}
\affiliation{Centro de Astrobiolog\'{\i}a (CAB), CSIC-INTA, Camino Bajo del Castillo s/n, E-28692 Villanueva de la Ca\~nada, Madrid, Spain}
\affiliation{Department of Physics, University of Oxford, Keble Road, Oxford OX1 3RH, UK}

\author{Chris Packham}
\affiliation{The University of Texas at San Antonio, One UTSA Circle, San Antonio, TX 78249, USA; lulu.zhang@utsa.edu; l.l.zhangastro@gmail.com}
\affiliation{National Astronomical Observatory of Japan, National Institutes of Natural Sciences (NINS), 2-21-1 Osawa, Mitaka, Tokyo 181-8588, Japan}

\author{Fergus R. Donnan}
\affiliation{Department of Physics, University of Oxford, Keble Road, Oxford OX1 3RH, UK}

\author{Dimitra Rigopoulou}
\affiliation{Department of Physics, University of Oxford, Keble Road, Oxford OX1 3RH, UK}
\affiliation{School of Sciences, European University Cyprus, Diogenes street, Engomi, 1516 Nicosia, Cyprus}

\author{Erin K. S. Hicks}
\affiliation{Department of Physics and Astronomy, University of Alaska Anchorage, Anchorage, AK 99508-4664, USA}
\affiliation{The University of Texas at San Antonio, One UTSA Circle, San Antonio, TX 78249, USA; lulu.zhang@utsa.edu; l.l.zhangastro@gmail.com}
\affiliation{Department of Physics, University of Alaska, Fairbanks, Alaska 99775-5920, USA}

\author{Ric I. Davies}
\affiliation{Max-Planck-Institut für extraterrestrische Physik, Postfach 1312, D-85741, Garching, Germany}

\author{Taro T. Shimizu}
\affiliation{Max-Planck-Institut für extraterrestrische Physik, Postfach 1312, D-85741, Garching, Germany}

\author{Almudena Alonso-Herrero}
\affiliation{Centro de Astrobiolog\'{\i}a (CAB), CSIC-INTA, Camino Bajo del Castillo s/n, E-28692 Villanueva de la Ca\~nada, Madrid, Spain}

\author{Cristina Ramos Almeida}
\affiliation{Instituto de Astrof{\'i}sica de Canarias, Calle V{\'i}a L{\'a}ctea, s/n, E-38205, La Laguna, Tenerife, Spain}
\affiliation{Departamento de Astrof{\'i}sica, Universidad de La Laguna, E-38206, La Laguna, Tenerife, Spain}

\author{Miguel Pereira-Santaella}
\affiliation{Instituto de F{\'i}sica Fundamental, CSIC, Calle Serrano 123, 28006 Madrid, Spain}

\author[0000-0001-5231-2645]{Claudio Ricci}
\affiliation{Instituto de Estudios Astrof{\'i}sicos, Facultad de Ingenier{\'i}a y Ciencias, Universidad Diego Portales, Av. Ej{\'e}rcito Libertador 441, Santiago, Chile}
\affiliation{Kavli Institute for Astronomy and Astrophysics, Peking University, Beijing 100871, People’s Republic of China}

\author{Andrew J. Bunker}
\affiliation{Department of Physics, University of Oxford, Keble Road, Oxford OX1 3RH, UK}

\author{Mason T. Leist}
\affiliation{The University of Texas at San Antonio, One UTSA Circle, San Antonio, TX 78249, USA; lulu.zhang@utsa.edu; l.l.zhangastro@gmail.com}

\author{David J. Rosario}
\affiliation{School of Mathematics, Statistics and Physics, Newcastle University, Newcastle upon Tyne, NE1 7RU, UK}

\author{Santiago Garc{\'i}a-Burillo}
\affiliation{Observatorio Astron{\'o}mico Nacional (OAN-IGN)-Observatorio de Madrid, Alfonso XII, 3, 28014, Madrid, Spain}

\author{Laura Hermosa Mu{\~n}oz}
\affiliation{Centro de Astrobiolog\'{\i}a (CAB), CSIC-INTA, Camino Bajo del Castillo s/n, E-28692 Villanueva de la Ca\~nada, Madrid, Spain}

\author{Francoise Combes}
\affiliation{LERMA, Observatoire de Paris, Coll{\`e}ge de France, PSL University, CNRS, Sorbonne University, Paris}

\author{Masatoshi Imanishi}
\affiliation{National Astronomical Observatory of Japan, National Institutes of Natural Sciences, 2-21-1 Osawa, Mitaka, Tokyo 181-8588, Japan}
\affiliation{Department of Astronomy, School of Science, Graduate University for Advanced Studies (SOKENDAI), Mitaka, Tokyo 181-8588, Japan}

\author[0000-0002-0690-8824]{Alvaro Labiano}
\affiliation{Telespazio UK for the European Space Agency (ESA), ESAC, Camino Bajo del Castillo s/n, 28692 Villanueva de la Ca{\~n}ada, Spain}

\author{Donaji Esparza-Arredondo}
\affiliation{Instituto de Astrof{\'i}sica de Canarias, Calle V{\'i}a L{\'a}ctea, s/n, E-38205, La Laguna, Tenerife, Spain}
\affiliation{Departamento de Astrof{\'i}sica, Universidad de La Laguna, E-38206, La Laguna, Tenerife, Spain}

\author{Enrica Bellocchi}
\affiliation{Departmento de F{\'i}sica de la Tierra y Astro{\'i}sica, Fac. de CC F{\'i}sicas, Universidad Complutense de Madrid, E-28040 Madrid, Spain}
\affiliation{Instituto de F{\'i}sica de Partículas y del Cosmos IPARCOS, Fac. CC 833 F{\'i}sicas, Universidad Complutense de Madrid, E-28040 Madrid, Spain}

\author{Anelise Audibert}
\affiliation{Instituto de Astrof{\'i}sica de Canarias, Calle V{\'i}a L{\'a}ctea, s/n, E-38205, La Laguna, Tenerife, Spain}
\affiliation{Departamento de Astrof{\'i}sica, Universidad de La Laguna, E-38206, La Laguna, Tenerife, Spain}

\author{Lindsay Fuller}
\affiliation{The University of Texas at San Antonio, One UTSA Circle, San Antonio, TX 78249, USA; lulu.zhang@utsa.edu; l.l.zhangastro@gmail.com}

\author{Omaira Gonz{\'a}lez-Mart{\'i}n}
\affiliation{Instituto de Radioastronom{\'i}a and Astrof{\'i}sica (IRyA-UNAM), 3-72 (Xangari), 8701, Morelia, Mexico}

\author{Sebastian H{\"o}nig}
\affiliation{School  of Physics \& Astronomy, University of Southampton, Hampshire SO17 1BJ, Southampton, UK}

\author{Takuma Izumi}
\affiliation{National Astronomical Observatory of Japan, National Institutes of Natural Sciences, 2-21-1 Osawa, Mitaka, Tokyo 181-8588, Japan}
\affiliation{Department of Astronomy, School of Science, Graduate University for Advanced Studies (SOKENDAI), Mitaka, Tokyo 181-8588, Japan}

\author[0000-0003-4209-639X]{Nancy A. Levenson}
\affiliation{Space Telescope Science Institute, 3700 San Martin Drive Baltimore, Maryland 21218, USA}

\author{Enrique L{\'o}pez-Rodríguez}
\affiliation{Department of Physics \& Astronomy, University of South Carolina, Columbia, SC 29208, USA}
\affiliation{Kavli Institute for Particle Astrophysics \& Cosmology (KIPAC), Stanford University, Stanford, CA 94305, USA}

\author{Daniel Rouan}
\affiliation{LESIA, Observatoire de Paris, Universit{\'e} PSL, CNRS, Sorbonne Universit{\'e}, Sorbonne Paris Cite{\'e}, 5 place Jules Janssen, 92195 Meudon, France}

\author{Marko Stalevski}
\affiliation{Astronomical Observatory, Volgina 7, 11060 Belgrade, Serbia}
\affiliation{Sterrenkundig Observatorium, Universiteit Gent, Krijgslaan 281-S9, Gent B-9000, Belgium}

\author{Martin J. Ward}
\affiliation{Centre for Extragalactic Astronomy, Durham University, South Road, Durham DH1 3LE, UK}



\begin{abstract}

We analyze JWST MIRI/MRS IFU observations of three Seyferts and showcase the intriguing polycyclic aromatic hydrocarbon (PAH) emission characteristics in regions of $\sim 500\,\rm pc$ scales over or around their active galactic nuclei (AGN). Combining the model predictions and the measurements of PAH features and other infrared emission lines, we find that the central regions containing a high fraction of neutral PAHs with small sizes, e.g., those in ESO137-G034, are in highly heated environments, due to collisional shock heating, with hard and moderately intense radiation fields. Such environments are proposed to be associated with inhibited growth or preferential erosion of PAHs, decreasing the average PAH size and the overall abundance of PAHs. We additionally find that the central regions containing a high fraction of ionized PAHs with large sizes, e.g., those in MCG-05-23-016, are likely experiencing severe photo-ionization because of the radiative effects from the radiative shock precursor besides the AGN. The severe photo-ionization can contribute to the ionization of all PAHs and further destruction of small PAHs. Overall, different Seyferts, even different regions in the same galaxy, e.g., those in NGC\,3081, can contain PAH populations of different properties. Specifically, Seyferts that exhibit similar PAH characteristics to ESO137-G034 and MCG-05-23-016 also tend to have similar emission line properties to them, suggesting that the explanations for PAH characteristics of ESO137-G034 and MCG-05-23-016 may also apply generally. These results have promising application in the era of JWST, especially in diagnosing different (i.e., radiative, and kinetic) AGN feedback modes.

\end{abstract}


\section{Introduction}

Polycyclic aromatic hydrocarbons (PAHs;  e.g., \citealt{Duley&Williams1981, Leger&Puget1984, Allamandola.etal.1985}; and see review \citealt{Tielens2008}) produce a series of  prominent emission features in the mid-infrared (MIR) spectra of star-forming galaxies and AGN. The main PAH features at 6.2, 7.7, 8.6, 11.3, and 12.7 $\mum$ together can account for up to 20\% of their total IR emission (\citealt{Smith.etal.2007, Xie.etal.2018}; and see review \citealt{Li2020}). PAH emission is proposed to be an effective estimator of the intensity of the ultraviolet (UV) radiation field, and hence the strength of recent star formation activity (e.g., \citealt{Rigopoulou.etal.1999, ForsterSchreiber.etal.2004, Peeters.etal.2004}). This is because PAH emission arises following the vibrational excitation of PAHs after absorbing a single UV photon primarily from young stars (e.g., \citealt{Allamandola.etal.1989, Tielens2005}), albeit partially from evolved stars (e.g., \citealt{Bendo.etal.2020, Zhang&Ho2023b}). Accordingly, PAH emission has been widely calibrated as an indicator of star formation rate for different galaxy environments (e.g., \citealt{Calzetti.etal.2005, Calzetti.etal.2007, Wu.etal.2005, Treyer.etal.2010, Shipley.etal.2016, Maragkoudakis.etal.2018, Xie&Ho2019, Belfiore.etal.2023, Ronayne.etal.2024}), and also as an indicator of molecular gas content (e.g., \citealt{Cortzen.etal.2019, Gao.etal.2019, Alonso-Herrero.etal.2020, Chown.etal.2021, Leroy.etal.2023, Whitcomb.etal.2023, Zhang&Ho2023a, Shivaei&Boogaard2024}).

From observational studies in the Spitzer era (see review \citealt{Li2020}), the relative intensity of individual PAH features has also been found to vary greatly across different galaxy environments, and such variability has not been fully explained in theory yet (e.g., \citealt{Genzel.etal.1998, Draine&Li2007, Farrah.etal.2007, Smith.etal.2007, Gordon.etal.2008, Kaneda.etal.2008, ODowd.etal.2009, Diamond-Stanic&Rieke2010, Hunt.etal.2010, Sales.etal.2010, Lebouteiller.etal.2011, Maragkoudakis.etal.2018, Maragkoudakis.etal.2022, Draine.etal.2021, Rigopoulou.etal.2021, Rigopoulou.etal.2024, Zhang.etal.2021, Zhang.etal.2022, Garcia-Bernete.etal.2022a, Garcia-Bernete.etal.2022b, Xie&Ho2022}). In particular, PAH features are generally weak and/or are diluted, especially for PAH features at short wavelengths (i.e., PAH 6.2, and 7.7 $\mum$ features), around AGN (e.g., \citealt{Smith.etal.2007, ODowd.etal.2009, Sales.etal.2010, Alonso-Herrero.etal.2014, RamosAlmeida.etal.2014, Garcia-Bernete.etal.2015}) and in low-metallicity star-forming regions (e.g., \citealt{Gordon.etal.2008, Hunt.etal.2010, Lebouteiller.etal.2011, Maragkoudakis.etal.2018}). In addition, while PAHs can survive in AGN periphery when with the shielding effect provided by molecular gas of high column densities (e.g., \citealt{Alonso-Herrero.etal.2014, Alonso-Herrero.etal.2020, Garcia-Bernete.etal.2022a}), such a trend, i.e., weak PAH emission, is more significant around AGN than in low-metallicity regions at given hardness of the radiation field (e.g., \citealt{Hunt.etal.2010, Lebouteiller.etal.2011}). This result hints that more than radiative effects, e.g., mechanical effects by shocks, are operating on PAHs around AGN (e.g., \citealt{Diamond-Stanic&Rieke2010, Garcia-Bernete.etal.2022a, Garcia-Bernete.etal.2024c, Zhang.etal.2022}).

Along with observational evidence, most of the above studies also provide specific explanations for the unique PAH characteristics around AGN. Therein, the selective erosion and destruction of smaller PAHs in the harsh environment around AGN are usually taken as the primary reason (e.g., \citealt{Smith.etal.2007, ODowd.etal.2009, Diamond-Stanic&Rieke2010, Sales.etal.2010, Garcia-Bernete.etal.2022a, Xie&Ho2022, Zhang.etal.2022, RamosAlmeida.etal.2023}). Moreover, such erosion and destruction are mostly attributed to the radiative effects of extreme-UV or X-ray photons in AGN periphery (\citealt{Aitken&Roche1985, Voit1992}), while the mechanical effects from low velocity shocks could also play a role under specific situations (e.g., \citealt{Diamond-Stanic&Rieke2010, Zhang.etal.2022, Donnan.etal.2023a}). Additionally, different obscuration levels of the environment (e.g., \citealt{Alonso-Herrero.etal.2014, Alonso-Herrero.etal.2020, Garcia-Bernete.etal.2022a, Garcia-Bernete.etal.2024c, Donnan.etal.2023b}) and different spectral energy distribution (SED) shapes of the radiation field (e.g., \citealt{ODowd.etal.2009, Draine.etal.2021, Rigopoulou.etal.2021, Donnelly.etal.2024}) that PAHs are immersed in can contribute to the unique PAH characteristics around AGN as well.

\begin{figure*}[t]
\center{\includegraphics[width=0.95\linewidth]{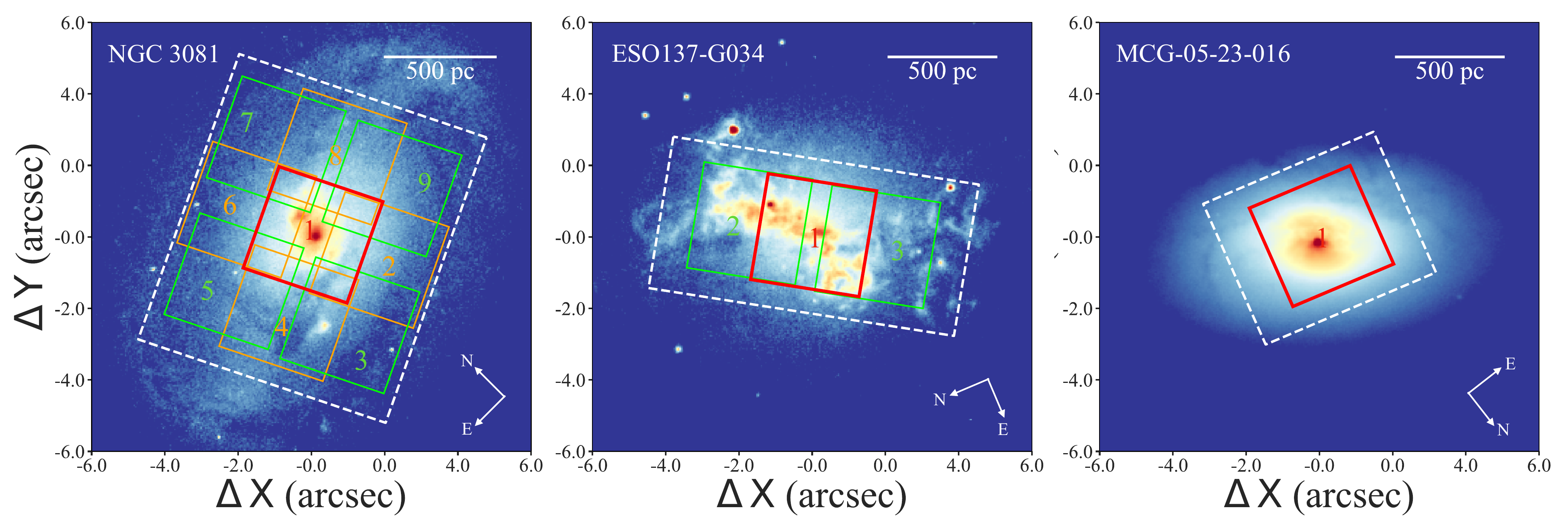}}
\caption{Illustration of coverages of the JWST/MRS observations (white dashed rectangles for channel 1 field of view) and relative positions of the $3\arcsec\times3\arcsec$ apertures (colored rectangles with corresponding numbers) of our targets, with HST/WFC3 F673N or F606W (MCG-05-23-016) band imaging as the background. A scale bar of 500 pc is in the top-right and a compass is in the bottom-right of each panel.}\label{coverage}
\end{figure*}

Although aforementioned mechanisms are able to explain the PAH characteristics of different AGN samples, a comprehensive understanding of the PAH characteristics under different conditions around AGN is still lacking. This is partly because these explanations essentially rely on the theoretical calculation of PAH emission and other emission line diagnostics tracing the underlying physics (e.g., [Ne~{\small III}] 15.5 $\mum$/[Ne~{\small II}]  12.8 $\mum$ as an indicator of radiation field hardness; \citealt{Thornley.etal.2000}), while the limited spatial and spectral resolutions of previous observations hamper the leverage of those powerful emission line diagnostics. With spectrographs of excellent spatial/spectral resolution, sensitivity, and wavelength coverage, the newly available James Webb Space Telescope (JWST; \citealt{Gardner.etal.2023, Rigby.etal.2023}) now provides an unprecedented opportunity to achieve a more comprehensive understanding of the PAH characteristics around AGN (e.g., \citealt{Garcia-Bernete.etal.2022b, Garcia-Bernete.etal.2024b, Garcia-Bernete.etal.2024c, Lai.etal.2022, Lai.etal.2023, Donnan.etal.2023a, Donnan.etal.2023b, Donnan.etal.2024, Zhang&Ho2023c}).

Leveraging capabilities of the Medium Resolution Spectrograph (MRS; \citealt{Wells.etal.2015, Labiano.etal.2021, Argyriou.etal.2023}) on the Mid-Infrared Instrument (MIRI; \citealt{Rieke.etal.2015, Wright.etal.2015, Wright.etal.2023}) onboard JWST, this letter showcases the diversity of PAH characteristics within the central kpc scale regions of three nearby Seyfert galaxies. The goal of this letter is combining the analysis of PAH features (Section~\ref{sec3}) and other emission line diagnostics (Section~\ref{sec4}), to shed light on the underlying mechanisms responsible for the varying PAH characteristics of the three targets (Section~\ref{sec5}).

\section{Data and Measurements}\label{sec2}

\subsection{Targets and Analysis}\label{sec2.1}

This letter is part of a series studying in total of six type 1.9/2 Seyferts with JWST MIRI/MRS integral field unit (IFU) spectral observations obtained by the JWST cycle 1 GO program (\#1670; PI: Shimizu, T. Taro). The six targets are all included in the Galactic Activity, Torus, and Outflow Survey (GATOS; \citealt{Garcia-Burillo.etal.2021, Garcia-Burillo.etal.2024, Alonso-Herrero.etal.2021, Garcia-Bernete.etal.2024a, Zhang.etal.2024})\footnote{\url{https://gatos.myportfolio.com}}. The full GATOS sample is a nearly complete selection of AGN with luminosities $L_{14-150\,\rm keV} > 10^{42}\,\rm erg\,s^{-1}$ at distances of $\sim10-40$ Mpc from the 70 Month Swift-BAT All sky Hard X-ray Survey (\citealt{Baumgartner.etal.2013}). This sample is also largely unbiased to obscuration/absorption even up to column densities of $N_{\rm H}\approx10^{24}\,\rm cm^{-2}$, from which the six targets of GO program \#1670 are further selected based on their outflow properties. A first analysis of the data set of this program, focussed on silicate features and water ice at small scales, is presented by \cite{Garcia-Bernete.etal.2024a}, followed by analyses focused on different aspects of one or more of the six targets (e.g., \citealt{Davies.etal.2024, Esparza-Arredondo.etal.2024, Garcia-Bernete.etal.2024c, Hermosa-Munoz.etal.2024, Zhang.etal.2024}, and Delany et al. in prep., Haidar et al. in prep.).

In particular, to ascertain the properties of PAH populations in the projected directions of AGN-driven outflows, \citeauthor{Garcia-Bernete.etal.2024c} (2024c; Paper~I hereafter) studied three targets with the strongest PAH emission and the most extended outflow regions among the six. Based on the spatially resolved analysis of PAH properties in different regions of the three targets, Paper~I revealed that the AGN affects not only the PAH population in the innermost regions but also in the extended outflow regions up to kpc scales. Specifically, whereas star-forming regions in the three targets are still located in the same zone of the diagram as the average values of star-forming galaxies, the outflow regions of these AGN occupy similar positions on the PAH diagrams as their innermost ($\sim 75\,\rm pc$) regions and AGN in the literature, which show a larger fraction of neutral PAH molecules.

In this letter, we focus on the remaining three targets, which exhibit relatively weak PAH emission without significantly extended outflow regions and cover a considerable range of AGN strength. The goal of this work is to shed light on the specific mechanisms that could be responsible for the variation of PAH properties around different AGN. To this end, this letter combines the analysis of both observational measurements and theoretical models of not only PAH features but also other infrared emission lines, which is among the first attempts of such work in the era of JWST. Given the relatively weak PAH emission of the three targets studied here, after some testing work, we choose to proceed the analysis with the MRS spectra extracted with a series of $3\arcsec\times3\arcsec$ apertures (see Figure~\ref{coverage}, $\sim$ 500 pc in physical scale for each, well beyond the nuclear regions studied by \citealt{Garcia-Bernete.etal.2024c}), rather than the spatially resolved MRS spectra. These apertures are simply selected to match the field of view of MRS IFU units.

\startlongtable
\setlength{\tabcolsep}{6pt}
\begin{deluxetable*}{cccccccc}
\tablecolumns{8}
\tablecaption{Properties of the Targets}
\tablehead{
\colhead{Galaxy} & \colhead{Type} & \colhead{$z$} & \colhead{$D_{L}$} & \colhead{log $N_{\rm H}$} & \colhead{log $L_{\rm bol}$} & \colhead{$\frac{L_{\rm bol}}{L_{\rm Edd}}$} & \colhead{$\dot{M}_{\rm out}$} \\
\colhead{(-)} & \colhead{(-)} & \colhead{(-)} & \colhead{(Mpc)} & \colhead{[$\rm cm^{-2}$]} & \colhead{[$\rm erg\,s^{-1}$]} & \colhead{(-)} & \colhead{($\rm M_{\odot}\,yr^{-1}$)} \\
\colhead{(1)} & \colhead{(2)} & \colhead{(3)} & \colhead{(4)} & \colhead{(5)} & \colhead{(6)} & \colhead{(7)} & \colhead{(8)}}
\startdata
NGC\,3081 & (R)SAB0/a(r) & 0.00798 & 34 & 23.9 & 44.1 & 0.02 & 0.03 \\
ESO137-G034 & SAB0/a & 0.00914 & 35 & 24.3 & 43.4 & 0.01& 0.33 \\
MCG-05-23-016 & S0 & 0.00849 & 35 & 22.2 & 44.3 & 0.06 & 0.04 \\
\enddata
\tablecomments{\footnotesize Column (1): Target name. Column (2): Target host type from NASA/IPAC Extragalactic Database (NED), with (r) and (R) indicating inner and outer ring, respectively. Column (3): Redshift taken from NED. Column (4): Luminosity distance taken from NED using redshift independent estimates or peculiar velocity corrections (\citealt{Theureau.etal.2007}). Column (5): Hydrogen column density based on modeling 0.3-150 keV X-ray spectrum (\citealt{Davies.etal.2015, Ricci.etal.2017}). Columns (6) \& (7): Bolometric AGN luminosity derived from intrinsic X-ray luminosity, and corresponding Eddington ratio (\citealt{Davies.etal.2015, Caglar.etal.2020}). Column (8): Ionized gas mass outflow rate within a $r = 0\farcs9$ aperture derived from the [Ne~{\scriptsize V}] 14.322 $\mum$\ emission line (\citealt{Zhang.etal.2024}).}
\label{tabinfo}
\end{deluxetable*}

Table~\ref{tabinfo} provides a brief summary of the three targets studied here, and more details about the sample and observations are presented in \cite{Garcia-Bernete.etal.2024a} and \cite{Zhang.etal.2024}. We primarily follow the standard JWST MIRI/MRS pipeline (release version 1.11.4) to reduce the raw data (\citealt{Labiano.etal.2016, Bushouse.etal.2023}), with the data reduction and extra steps detailed in \cite{Garcia-Bernete.etal.2024a}. We use the same configuration (calibration context 1130) of the pipeline stages as in \cite{Garcia-Bernete.etal.2022a} and \cite{Pereira-Santaella.etal.2022}. We also apply an extra JWST pipeline step (i.e., $\tt residual\_fringe$) not implemented in the standard JWST pipeline to correct the low-frequency fringe residuals (\citealt{Law.etal.2023}), which remain after applying the standard fringe removal and could have a significant influence on weak spectral features (\citealt{Argyriou.etal.2020, Gasman.etal.2023}). Finally, we apply another extra step to mask some hot and cold pixels that are not identified by the standard pipeline before creating the 3D spectra data cubes.

\subsection{PAH and Emission Line Measurements}\label{sec2.2}
The following analysis is mainly based on the first three MRS channels (i.e., channel 1, 2, and 3; $\lambda \approx 5-18\,\mum$). These three channels cover most commonly used PAH features and emission lines relevant for this study except for the [O~{\footnotesize IV}] 25.89~$\mum$ emission line, which was extracted from the MRS channel 4 spectral data cube. All the slices from the spectral data cubes of the first three channels are convolved to the same angular resolution before the extraction of spectra and measurements for all targets (see more details in \citealt{Zhang.etal.2024}). The extracted spectra are stitched to the channel 3 spectra, based on the median values of the overlapping portion of different band spectra. An extra 10~\% uncertainty according to the scaling factors is included in quadrature sum during the PAH decomposition as detailed below.

\begin{figure*}[!ht]
\center{\includegraphics[width=0.675\linewidth]{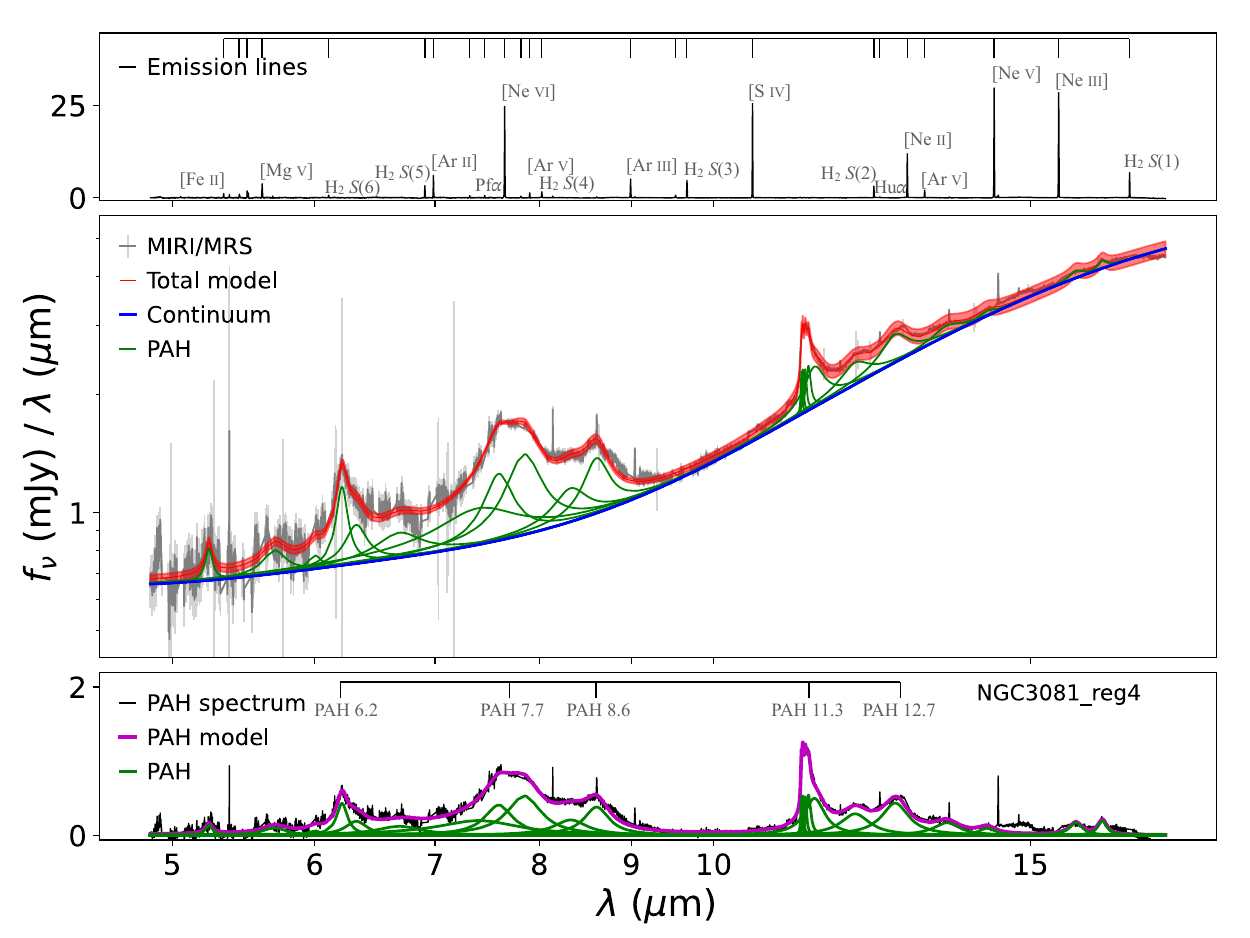}}
\caption{Top panel: The residual emission line spectrum after subtracting the best-fit model (i.e., the red curve in the middle panel) from the MIRI/MRS spectrum. From the left to the right, the short lines in the top indicate the positions of [Fe~{\footnotesize II}] 5.34, [Fe~{\footnotesize VIII}] 5.45, [Mg~{\footnotesize VII}] 5.50, [Mg~{\footnotesize V}] 5.61, ${\rm H}_{2}\, S(6)$, ${\rm H}_{2}\, S(5)$, [Ar~{\footnotesize II}] 6.985, [Na~{\footnotesize III}] 7.32, Pf$\alpha$, [Ne~{\footnotesize VI}] 7.65, [Fe~{\footnotesize VII}] 7.815, [Ar~{\footnotesize V}] 7.90, ${\rm H}_{2}\, S(4)$, [Ar~{\footnotesize III}] 8.99, [Fe~{\footnotesize VII}] 9.53, ${\rm H}_{2}\, S(3)$, [S~{\footnotesize IV}] 10.51, ${\rm H}_{2}\, S(2)$, Hu$\alpha$, [Ne~{\footnotesize II}] 12.81, [Ar~{\footnotesize V}] 13.10, [Ne~{\footnotesize V}] 14.32, [Ne~{\footnotesize III}] 15.555, and ${\rm H}_{2}\, S(1)$ emission lines, respectively. Middle panel: Illustration of the multi-component fitting for PAH measurements (see Section~\ref{sec2.2}). The gray curve is the observed MIRI/MRS spectrum with emission lines masked. The red curve gives the best-fit model with the shaded region indicating the corresponding 1 $\sigma$ posterior distribution. The blue curve is the summation of all continuum components (i.e., stellar and dust continuum), and these green curves are Drude profiles representing individual PAH features. Bottom panel: The residual PAH spectrum (i.e., the black curve) after subtracting all continuum components (i.e., the blue curve in the middle panel) from the emission line masked MIRI/MRS spectrum. Same as in the middle panel, these green curves are Drude profiles representing individual PAH features, and the magenta curve is the summation of all green curves, i.e., the modeled PAH spectrum. From the left to the right, the short lines in the top indicate the positions of PAH complexes around 6.2, 7.7, 8.6, 11.3, and 12.7 $\mum$, respectively. Note that all x-axes and the y-axis of the middle panel are in logarithmic scale, while y-axes of the top and bottom panels are in linear scale. The same plots for all apertures studied here are available from Figure~\ref{Fit_Demos} in Appendix.}\label{Fit_Demo}
\end{figure*}

The PAH features are decomposed and measured through multi-component fitting from the first three channel spectra extracted with the $3\arcsec\times3\arcsec$ apertures after masking all emission lines (see Figure~\ref{Fit_Demo}). The multi-component model for the fitting consists of a series of Drude profiles for individual PAH features, nine modified blackbodies of fixed temperatures and a black body of 5000 $\rm K$ for the underlying continuum, all subject to dust attenuation by foreground extinction. Specifically, the Drude profiles with fixed widths are primarily adopted from \cite{Draine&Li2007} with some updates according to \cite{Donnan.etal.2023a} for fitting MRS spectra. The temperatures of the modified blackbodies are 35, 40, 50, 65, 90, 135, 200, 300, and 500 $\rm K$, respectively, in accordance with previous work (i.e., \citealt{Smith.etal.2007, Donnan.etal.2023a}). We have checked that adopting modified blackbodies of different temperatures among this range and including more modified blackbodies of higher temperatures does not change much the measurement of PAH features and therefore does not affect our conclusions. The 5000\,$\rm K$ black body is to mimic the contribution of stellar continuum. The foreground extinction extinction (i.e., $e^{-\tau_{\lambda}}$) is fitted based on the optical depth curve (i.e., $\tau_{\lambda}$) measured by \cite{Garcia-Bernete.etal.2024a} for each target with the scaling factor of the optical depth curve as a free parameter.

\begin{figure*}[!ht]
\center{\includegraphics[width=0.65\linewidth]{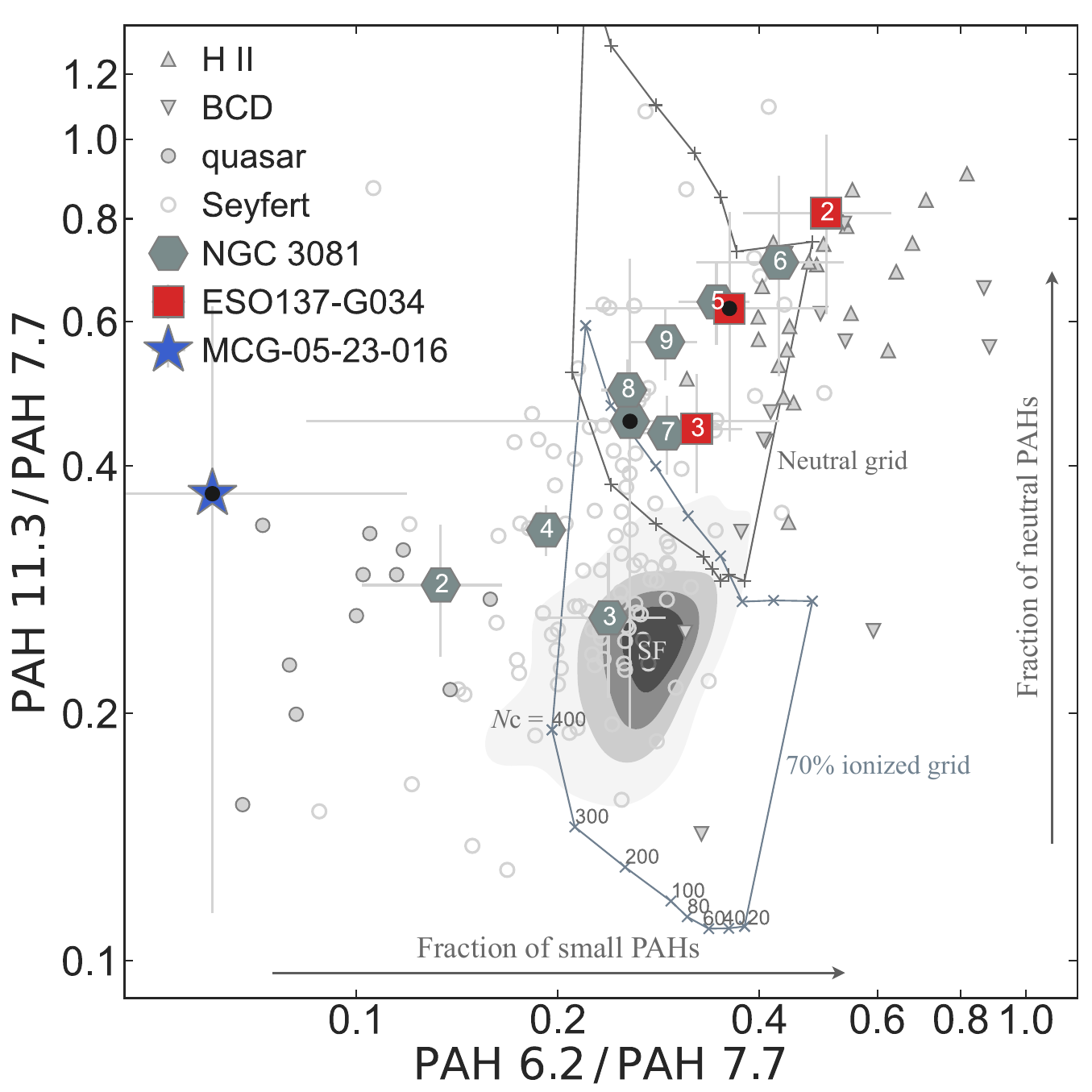}}
\caption{Diagnostic diagram of PAH band ratios 11.3\,$\mum$/7.7\,$\mum$ versus 6.2\,$\mum$/7.7\,$\mum$ for the $3\arcsec\times3\arcsec$ ($\sim 500\,\rm pc$) apertures as illustrated in Figure~\ref{coverage}. The red rectangles, blue star, slategray hexagons represent the measurements of apertures in ESO137-G034, MCG-05-23-016, and NGC\,3081, respectively, where points with a black dot pertain to the innermost aperture (i.e., aperture 1) of each target and other apertures are number-coded the same as in Figure~\ref{coverage}. The gray contours marked by ``SF'' represent the PAH band ratio distribution of 185 spatially resolved spaxels in 29 star-forming galaxies measured by \cite{Zhang.etal.2022}. The open small circles indicate the PAH band ratios of 83 Seyferts with the measurements of all the three PAH features available from the literature (i.e., \citealt{Smith.etal.2007, ODowd.etal.2009, Diamond-Stanic&Rieke2010, Gallimore.etal.2010, Sales.etal.2010}; PAH measurements of upper limits are excluded). The filled small circles indicate the PAH band ratios of 11 low-redshift quasars with all the three PAH features detected by \cite{Xie&Ho2022}. The upward and downward small triangles, respectively, indicate the PAH band ratios of 29 giant H~{\footnotesize II} regions in 3 local group galaxies measured by \cite{Lebouteiller.etal.2011} and 14 blue compact dwarf galaxies (BCDs) measured by \cite{Hunt.etal.2010} and \cite{Lebouteiller.etal.2011}.  The gray grids represent model predictions of PAH band ratios by \cite{Rigopoulou.etal.2024}, respectively, for neutral (top grid, marked by `$+$') and 70\% ionized (bottom grid, marked by `$\times$') PAHs of different sizes (i.e., with carbon number $N_{\rm C} = 20 - 400$ from the right boundary to the left boundary of each grid), in the interstellar radiation field (ISRF; the top boundary of each grid) and the $10^3\times$ ISRF (the bottom boundary of each grid).}\label{PAHratio}
\end{figure*}

The fitting is implemented based on the Bayesian Markov Chain Monte Carlo procedure $\tt emcee$ in the $\tt Python$ environment, with the median and standard deviation of the posterior distribution of each best-fit parameter taken as the final estimate and corresponding uncertainty. The following discussion focuses on the most prominent PAH features at 6.2, 7.7, and 11.3\,$\mum$ (see Table~\ref{tabPAHs} in Appendix). We have further checked the robustness of this multi-component fitting procedure in decomposing the continuum-dominant spectra, especially for that of MCG-05-23-016. Specifically, we generated a series of mock spectra combining the best-fit PAH spectra of MCG-05-23-016 with the nuclear dust continuum scaled by a different factor (i.e., $\times\,0.5-2.0$) and artificially changed extinction (i.e., $\times\,0.2-1.5$), plus corresponding noise. We then performed the same multi-component fitting for these mock spectra. The measured flux of PAH features at 6.2, 7.7, and 11.3\,$\mum$ for these mock spectra are consistent with the best-fit PAH flux of MCG-05-23-016 within 8\,\%. This value is far below the uncertainty of the best-fit PAH flux of MCG-05-23-016 obtained from the MCMC fitting procedure, thus indicating that the multi-component fitting procedure is robust in decomposing the continuum-dominant spectra.

Apart from PAH features, MIR ionic and molecular emission lines also provide valuable diagnostics of the physical conditions around AGN (e.g., \citealt{Pereira-Santaella.etal.2010, Pereira-Santaella.etal.2017, Sajina.etal.2022, Feltre.etal.2023}). The following analysis also involves six ionized emission lines, [Ne~{\footnotesize II}] 12.814~$\mum$, [Ne~{\footnotesize III}] 15.555~$\mum$, [Ne~{\footnotesize V}] 14.322~$\mum$, [S~{\footnotesize IV}] 10.511~$\mum$, [O~{\footnotesize IV}] 25.89~$\mum$, and [Fe~{\footnotesize II}] 5.34~$\mum$ (see Table~\ref{tabLines}), and eight hydrogen emission lines, ${\rm H}_{2}\, S(1) - {\rm H}_{2}\, S(6)$, Pf$\alpha$, and Hu$\alpha$ (see Table~\ref{tabH2s}). We adopt the same strategy as in \cite{Zhang.etal.2024}, which explored ionized gas outflows in central kiloparsec regions of the six Seyferts, for the fitting of the ionized and hydrogen emission lines. In brief, we fit these emission lines individually using a single- and then a double-Gaussian profiles, plus a local linear continuum based on the Levenberg-Marquardt least-squares minimization algorithm. Specifically, we have checked that for apertures studied here, except for [Fe~{\footnotesize II}], a double-Gaussian profile is required for the fitting of all ionized emission lines involved here because of the potential outflow components therein. Meanwhile, a single-Gaussian profile is good enough for the fitting of hydrogen emission lines in most cases. To get more robust statistics, the spectrum containing each emission line is perturbed with a random noise at the uncertainty level and then the fitting is repeated 100 times. The median and standard deviation of those 100 fits are taken for the final flux estimate and corresponding uncertainty of each emission line, respectively. Additionally, all the flux measurements of these emission lines are corrected for dust extinction according to the extinction strength derived from the multi-component full-spectrum fitting described above.

\section{PAH Characteristics of Central Apertures in Target Seyferts}\label{sec3}

Figure~\ref{PAHratio} shows the most widely used PAH diagram (i.e., 11.3\,$\mum$/7.7\,$\mum$ versus 6.2\,$\mum$/7.7\,$\mum$) with the PAH measurements for all $3\arcsec\times3\arcsec$ ($\sim$ 500 pc) apertures in the three targets, as well as the PAH measurements retrieved from previous work for the control samples. We find from the PAH diagram that positions occupied by apertures in the three targets are overall different from those occupied by spatially resolved spaxels in star-forming (SF) galaxies. 

Compared to the SF spaxels, the apertures of ESO137-G034 overall exhibit the much larger PAH 11.3\,$\mum$/7.7\,$\mum$ band ratios and slightly larger PAH 6.2\,$\mum$/7.7\,$\mum$ band ratios, while the aperture of MCG-05-23-016 exhibits the much smaller PAH 6.2\,$\mum$/7.7\,$\mum$ band ratio and the moderate PAH 11.3\,$\mum$/7.7\,$\mum$ band ratio. Moreover, the apertures of ESO137-G034 and MCG-05-23-016 on the PAH diagram are close to the loci of low metallicity star-forming systems (i.e., H~{\footnotesize II} regions and BCDs) and low redshift quasars, respectively. We will specifically discuss the similarity and difference between these systems in Section~\ref{sec4} to shed more light on the physical mechanisms that could be responsible for their PAH characteristics. Meanwhile, the apertures of NGC\,3081 exhibit the PAH band ratios of a wide distribution, overlapping with those of a large sample of Seyfert galaxies studied in the era of Spitzer. Overall, apertures of targets studied here have a wide distribution of PAH band ratios, covering those of most Seyferts, as well as those of the outflow regions and the innermost regions of the three targets studied in Paper~I.

Aperture 3 of NGC\,3081 has the PAH band ratios most similar to those of SF spaxels, which likely is due to the contribution of a SF blob therein (\citealt{Ma.etal.2021}). Similarly, the PAH band ratios of Seyferts that are overlapped with SF regions on the PAH diagram can also be explained by the contribution of extended star forming regions around their active nuclei (e.g., \citealt{Garcia-Bernete.etal.2022a, Zhang&Ho2023c}). The innermost apertures of the three targets have the PAH band ratios with large uncertainties, which are resulted form the uncertainties of their PAH measurements due to the high continuum levels of their spectra. Despite the large uncertainties, their PAH band ratios are physically reasonable considering other emission line diagnostics as detailed in Section~\ref{sec4}, especially for ESO137-G034 and MCG-05-23-016, which have unusual PAH band ratios compared to most Seyferts.


As depicted by the model grids in Figure~\ref{PAHratio}, the different ionization states and size distributions of PAHs can basically explain the distinct PAH characteristics of apertures in the three targets. Specifically, neutral PAHs produce stronger C-H modes responsible for the 3.3 and 11.3\,$\mum$ PAH features, while cationic PAHs produce more C-C modes emitting the $6-9$ $\mum$ PAH features (i.e., higher PAH 11.3\,$\mum$/7.7\,$\mum$ ratio for neutral PAHs). Meanwhile, under given excitation conditions, smaller PAHs, because of their low heat capacity, reach higher levels of vibrational excitation and hence radiate at shorter, more energetic wavelengths. Moreover, ionized PAHs, especially the small ones, are also more vulnerable to photo-destruction compared to the neutral ones in harsh environments (e.g., \citealt{Allain.etal.1996, Holm.etal.2011}). Partially destroyed PAHs with open/irregular structures and more hydrogen atoms (e.g., catacondensed PAHs) could exhibit higher PAH 11.3\,$\mum$/7.7\,$\mum$ (\citealt{Li2020}), while this result is still on debate as \cite{Rigopoulou.etal.2024} recently found dehydrogenation has little influence on the observed inter-band relations of PAHs.


According to the model predictions by \cite{Rigopoulou.etal.2024}, i.e., the gray grids in Figure~\ref{PAHratio}, the apertures in NGC\,3081 and ESO137-G034 with PAH $11.3\,\mum/7.7\,\mum > 0.4$ contain more neutral PAHs with the carbon number, hereafter $N_{\rm C}$, $\lesssim 300$. As noted, a high fraction of neutral PAHs can survive around AGN with the shielding effect provided by molecular gas of high column densities (e.g., \citealt{Alonso-Herrero.etal.2014, Alonso-Herrero.etal.2020, Garcia-Bernete.etal.2022a}). For these apertures with a high fraction of neutral PAHs, some of them, especially the three apertures of ESO137-G034, even contain a quite large fraction of PAHs with $N_{\rm C} \lesssim 100$, as seen in some low metallicity star-forming systems (i.e., H~{\small II} regions and BCDs as the upward and downward small gray triangles in Figure~\ref{PAHratio}; \citealt{Hunt.etal.2010, Lebouteiller.etal.2011}). This is a surprising result as smaller PAHs should be more easily affected under such harsh environments. About this point, we will have further discussion in Section~\ref{sec5}.

\begin{figure*}[!ht]
\center{\includegraphics[width=1\linewidth]{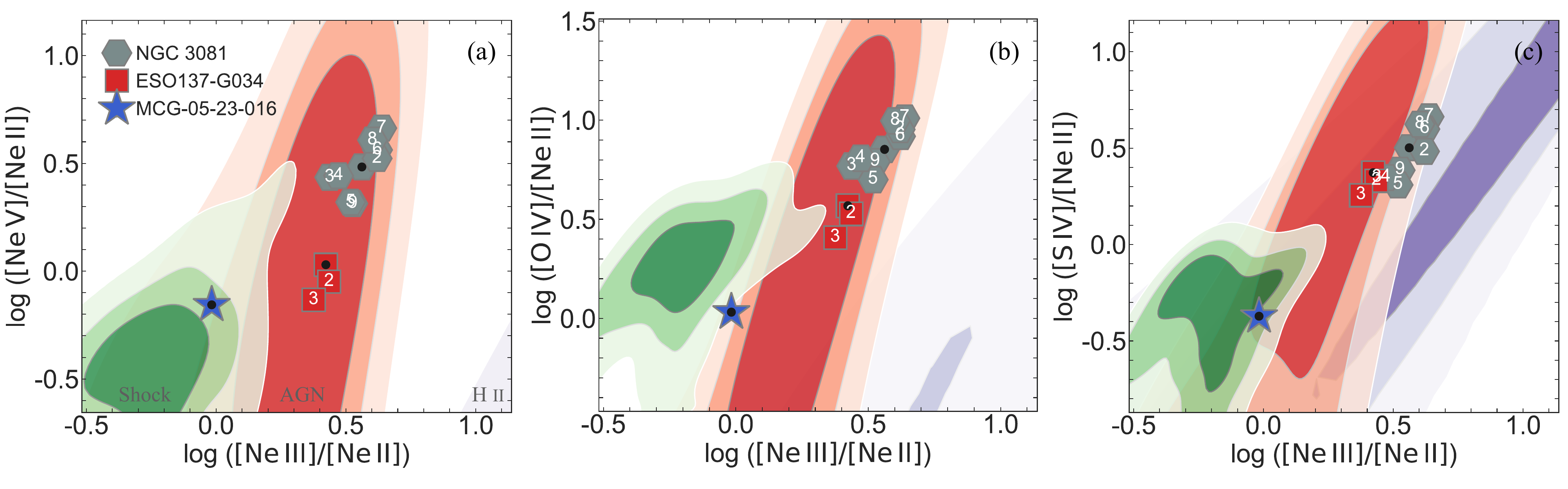}}
\caption{Diagnostic diagrams of ionized emission line ratios as (a) [Ne~{\footnotesize V}]/[Ne~{\footnotesize II}] versus [Ne~{\footnotesize III}]/[Ne~{\footnotesize II}], (b) [O~{\footnotesize IV}]/[Ne~{\footnotesize II}] versus [Ne~{\footnotesize III}]/[Ne~{\footnotesize II}], and (c) [S~{\footnotesize IV}]/[Ne~{\footnotesize II}] versus [Ne~{\footnotesize III}]/[Ne~{\footnotesize II}] for apertures in the three targets with markers the same as in Figure~\ref{PAHratio}. The reddish, greenish, and purplish contours in each panel correspond to the model calculations for AGN, fast radiative shocks (including the shock precursor), and H~{\footnotesize II} regions, respectively, with each contour containing 30\%, 60\%, and 90\% of the model results from the inside to the outside, respectively.}\label{Lineratio}
\end{figure*}

In addition, the leftmost three apertures of the three targets in Figure~\ref{PAHratio} (PAH $6.2\,\mum/7.7\,\mum \lesssim 0.2$) in theory contain more ionized PAHs of large sizes with $N_{\rm C} \gtrsim 400$, as also seen in some low-redshift quasars (i.e., small gray circles in Figure~\ref{PAHratio}; \citealt{Xie&Ho2022}). In particular, the observed PAH band ratios of MCG-05-23-016, although with large uncertainties, are significantly beyond the model predictions by \cite{Rigopoulou.etal.2024} for PAHs with $N_{\rm C} $ up to 400, and are consistent with the theoretical calculation by \cite{Draine&Li2001} for ionized PAHs with $N_{\rm C}$ more than 1000. There should be some unique processes (as will be discussed in Section~\ref{sec4}~\&~\ref{sec5}) that lead to the presence of more ionized PAHs with large sizes in the central region of MCG-05-23-016. 


\section{Emission Line Diagnostics of Central Apertures in Target Seyferts}\label{sec4}

As noted, MIR ionic and molecular emission lines covered by JWST MIRI/MRS spectra provide valuable diagnostics, which will help identify the underlying mechanisms resulting in the varying PAH characteristics around the AGN studied here. Figure~\ref{Lineratio} presents three diagnostic diagrams consisting of some mostly used MIR ionized emission line ratios (e.g., \citealt{Dale.etal.2006, Armus.etal.2007, Pereira-Santaella.etal.2010, Feltre.etal.2023}), and the distributions of these line ratios derived from theoretical models of AGN, fast radiative shocks, and H~{\small II} regions. The H~{\small II} and AGN models are calculated using {\tt C{\scriptsize LOUDY}} (\citealt{Ferland.etal.2017, Chatzikos.etal.2023}) by \cite{Morisset.etal.2015} and \cite{Pereira-Santaella.etal.2024}, respectively. The fast radiative shock models (including the shock precursor, with the shock velocity $v \approx 100 - 1000\,{\rm km\,s^{-1}}$, the pre-shock density $n \approx 10 - 10^{4}\,{\rm cm^{-3}}$, the transverse magnetic field $B \approx 0.1 - 3\,{\rm\mu G}$, and the metallicity $Z \approx 0.3 - 2.5\,Z_{\odot}$) are calculated using the {\tt MAPPINGS~V} (\citealt{Sutherland&Dopita2017}) by \cite{Alarie&Morisset2019}.\footnote{The SF and shock models are publicly available from the \href{https://sites.google.com/site/mexicanmillionmodels/home}{3MdB} and \href{http://3mdb.astro.unam.mx:3686/}{3MdBs} databases under the ref=`BOND\_2' and ref=`Gutkin16', respectively.}

We find that although the three apertures in ESO137-G034 exhibit the ionized emission line ratios consistent with the results obtained by AGN models, their highest ionization potential lines (e.g., [Ne~{\footnotesize V}] with the ionization potential, i.e., IP, of 97.1 eV) are remarkably weak compared to other high ionization potential lines (e.g., [O~{\footnotesize IV}] and [S~{\footnotesize IV}] with the IP of 54.9 eV and 34.8 eV, respectively). For reference, the apertures in NGC\,3081, which are also dominated by AGN excitation, exhibit overall strong emission of all the three high ionization potential lines. Specifically, AGN models calculated with incident radiation SEDs of lower Eddington ratios and smaller ionization parameters, i.e., $U$ as the indicator of ionization field intensity, exhibit lower [Ne~{\footnotesize V}]/[Ne~{\footnotesize II}] ratios (\citealt{Pereira-Santaella.etal.2024}). Accordingly, the lower [Ne~{\footnotesize V}]/[Ne~{\footnotesize II}] ratios reflect the weak accretion nature and moderately intense radiation field of ESO137-G034. Additionally, the radiation filed therein is hard given the large values of the [Ne~{\footnotesize III}]/[Ne~{\footnotesize II}] ratio, which is a canonical indicator of the radiation filed hardness (e.g., \citealt{Thornley.etal.2000, Groves.etal.2006}). Meanwhile, we find that the central region of MCG-05-23-016 is likely also affected by fast radiative shocks in addition to AGN excitation. However, the emission lines from the central aperture of MCG-05-23-016 do not show the significant broadening as expected for highly shocked regions. This suggests that the fast radiative shocks in MCG-05-23-016 are more likely playing a role in the radiative rather than kinetic manner as detailed later. 

\begin{figure}[t]
\center{\includegraphics[width=0.9\linewidth]{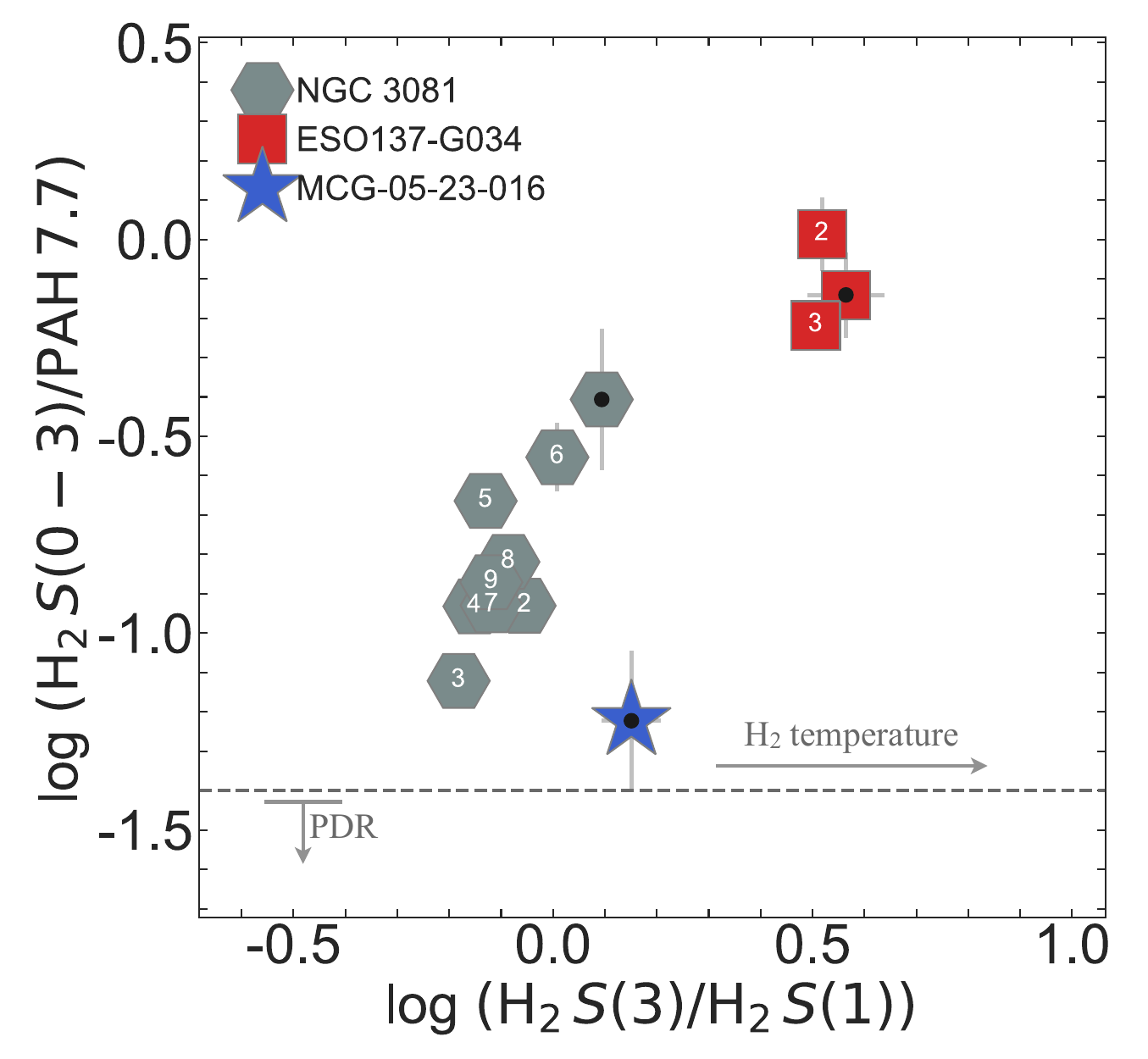}}
\caption{Diagnostic diagram of ${\rm H_{2}}\,S(0-3)/{\rm PAH\,7.7}$ and ${\rm H_{2}}\,S(3)/{\rm H_{2}}\,S(1)$ ratios for apertures in the three targets with markers the same as in Figure~\ref{PAHratio}. The horizontal dashed line shows the upper limit of ${\rm H_{2}}\,S(0-3)/{\rm PAH\,7.7}$  ratio given by the PDR models adopted by \cite{Guillard.etal.2012}, and the horizontal arrow indicates the direction of increasing H$_{2}$ temperature.}\label{LineratioII}
\end{figure}

Previous studies found that collisional (i.e., kinetic) shock heating associated with the AGN can contribute to an excess of infrared molecular hydrogen emission, especially relative to PAH emission (e.g., \citealt{Roussel.etal.2007, Ogle.etal.2010, Guillard.etal.2012}). Accordingly, we present in Figure~\ref{LineratioII} the diagnostic diagram of ${\rm H_{2}}\,S(0-3)/{\rm PAH\,7.7}$ and ${\rm H_{2}}\,S(3)/{\rm H_{2}}\,S(1)$ ratios, the latter of which provides a good proxy of different gas temperatures with the ${\rm H_{2}}\,S(3)/{\rm H_{2}}\,S(1)$ ratio ranging from $\sim 0.6 - 3$ for gas temperatures from $\sim 150 - 400\,\rm K$ (\citealt{Turner.etal.1977}). Note that ${\rm H_{2}}\,S(0-3)$ is the summation of ${\rm H_{2}}\,S(0)$ to ${\rm H_{2}}\,S(3)$, which is converted from the summation of ${\rm H_{2}}\,S(1)$ to ${\rm H_{2}}\,S(5)$ by dividing by a factor of 0.9 following \cite{Lai.etal.2022} based on the results of \cite{Habart.etal.2011}. We find that the ${\rm H_{2}}\,S(0-3)/{\rm PAH\,7.7}$ ratios for all apertures are greater than the theoretical threshold that can be generated by photo-dissociation regions, i.e., ${\rm log}\,({\rm H_{2}}\,S(0-3)/{\rm PAH\,7.7}) = -1.4$ (\citealt{Guillard.etal.2012}). Among apertures in the three targets, the apertures of ESO137-G034 exhibit the most significant excess, while the aperture of MCG-05-23-016 only exhibits slight excess of infrared H$_2$ emission. The relatively lower ${\rm H_{2}}\,S(0-3)/{\rm PAH\,7.7}$ ratio in the central region of MCG-05-23-016 is intrinsically due to weak H$_2$ emission therein, where Pf$\alpha$ and Hu$\alpha$ emission is in fact not weak relative to PAH emission (see Figure~\ref{H_PAH} in Appendix). 

Related to this result, we find that [Fe~{\small II}]/Pf$\alpha$ and [Fe~{\small II}]/Hu$\alpha$ ratios, as another empirical diagnostics of collisional shock heating, of the three targets exhibit the same trend as their ${\rm H_{2}}\,S(0-3)/{\rm PAH\,7.7}$ ratios. Namely, the apertures of ESO137-G034 and MCG-05-23-016 exhibit the highest and lowest [Fe~{\small II}]/Pf$\alpha$ as well as [Fe~{\small II}]/Hu$\alpha$ ratios, respectively, while the apertures of NGC\,3081 exhibits the values between those of ESO137-G034 and MCG-05-23-016 (see Figure~\ref{H_Fe}). In fact, the intrinsic [Fe~{\small II}] emission of MCG-05-23-016 is not weak, and its value is roughly the same as that of the innermost aperture of NGC\,3081 and greater than those of other apertures of NGC\,3081 (see Table~\ref{tabLines}). The aperture of MCG-05-23-016 exhibits the relatively weak evidence of collisional shock heating, but have the strongest [Ne~{\footnotesize II}] emission and the highest Pf$\alpha$/${\rm H_{2}}\,S(0-3)$ and Hu$\alpha$/${\rm H_{2}}\,S(0-3)$ ratios among all the 13 apertures of the three targets (see Figure~\ref{H_PAH}). Meanwhile, the three apertures of ESO137-G034 have the highest temperatures according to their ${\rm H_{2}}\,S(3)/{\rm H_{2}}\,S(1)$ ratios, consistent with its strongest signature of collisional shock heating. These results, combining the findings from Figure~\ref{Lineratio}, indicate that the central region of MCG-05-23-016 is more likely dominated by photo-ionization from the (fast radiative) shock precursor in addition to the AGN, i.e., the radiative mode feedback, rather than the collisional shock excitation, i.e., the kinetic mode feedback, while the central region of ESO137-G034 is highly heated due to the collisional shock excitation (see further discussion in Section~\ref{sec5}). 

\begin{figure}[t]
\center{\includegraphics[width=0.9\linewidth]{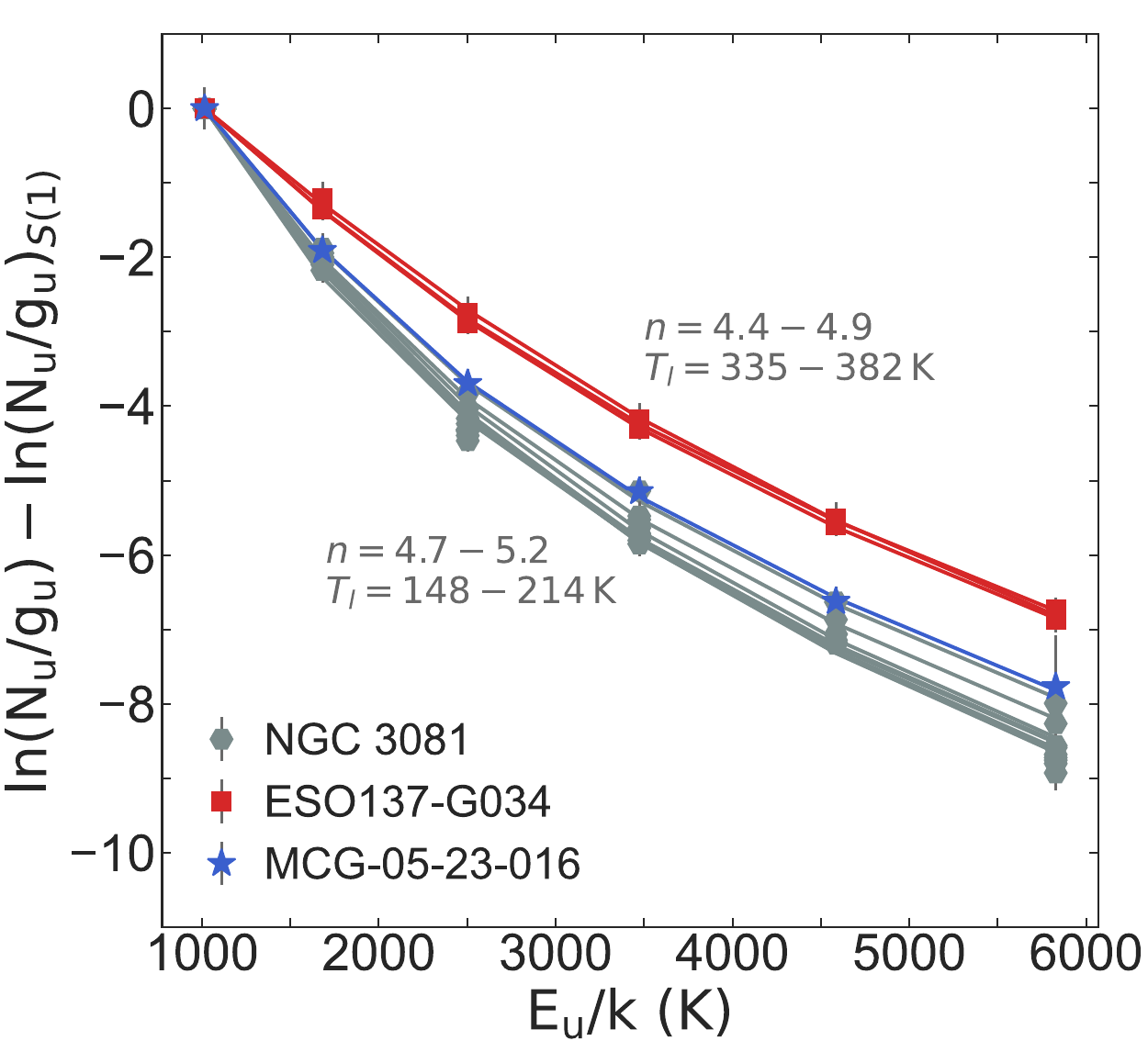}}
\caption{Fitting results of H$_2$ rotational emission lines assuming a continuous power-law distribution of temperature $T$ for apertures in the three targets with markers the same as in Figure~\ref{PAHratio}. The ranges of the best-fit temperature distribution index $n$ and low end temperature $T_{l}$ for ESO137-G034 and the other two targets are indicated above and below the fitting curves, respectively.}\label{H2_curves}
\end{figure}

We further fit the total flux of six hydrogen rotational lines, i.e., ${\rm H}_{2}\, S(1) - {\rm H}_{2}\, S(6)$, to provide a more quantitative estimate of the gas temperature in central regions of the three targets. Specifically, the flux of H$_2$ rotational lines at energy level $J$ is related to the column density $N(J+2)$ as $N(J+2) \propto \frac{F(J)\lambda}{A}$, with the $N(J)$ can be fitted as $N(J) \propto \int_{T_{l}}^{T_{u}}\frac{g(J)}{Z(T)} \, e^{-E(J)/kT}\ T^{-n}dT$, assuming a more realistic power-law distribution of the temperature $T$ (\citealt{Togi&Smith2016, Appleton.etal.2017}). We fit for the low end temperature ($T_{l}$) and the power-law index ($n$) of the temperature distribution with a fixed high end temperature $T_{u} = 2000\,{\rm K}$ following \cite{Togi&Smith2016}. As shown in Figure~\ref{H2_curves}, we obtain the highest $T_{l}$ (i.e., $\sim 335 - 382\,{\rm K}$) for the three apertures of ESO137-G034 among all the 13 apertures, while for the other apertures the $T_{l}$ is lower (i.e., $\sim 148 - 214\,{\rm K}$). Moreover, the three apertures of ESO137-G034 have overall smaller $n$ values ($\sim 4.4 - 4.9$) compared to the other apertures ($\sim 4.7 - 5.2$), indicating an even higher fraction of hot gas in the central region of ESO137-G034.

The fitted power-law index $n$ and low end temperature $T_{l}$, as well as the derived mass and column density of H$_{2}$ at $T > 200\,\rm K$ based on the best-fit result for each aperture are listed in Table~\ref{tabH2curves} in Appendix. We obtain from the best-fit results another important information that the central aperture of MCG-05-23-016 has the lowest column density (i.e., $10^{20.35}\,{\rm cm^{-2}}$) of H$_{2}$ at $T > 200\,\rm K$ among the three innermost apertures of the three targets. This result is consistent with the lowest column density (i.e., $10^{22.2}\,{\rm cm^{-2}}$) of total H in MCG-05-23-016 among the three targets (see Table~\ref{tabinfo}).

\section{Discussion: Underlying Mechanisms Responsible for the PAH Characteristics}\label{sec5}

Combining the emission line diagnostics as detailed in Section~\ref{sec4}, we can now discuss the underlying physics that might contribute to the systematically different PAH characteristics of the three targets studied here. 

The hard but moderately intense radiation fields and high temperatures of the three apertures in ESO137-G034 are qualitatively similar to low metallicity SF systems. Recently, \cite{Whitcomb.etal.2024} found a steep decline in the strength of long wavelength PAH features below solar metallicity, with the short wavelength PAH features carrying an increasingly large fraction of radiation energy. With the help of newly developed grain models, \cite{Whitcomb.etal.2024} found the data are consistent with an evolving grain size distribution that shifts to smaller sizes as metallicity declines. They further attributed such a shift to the inhibited grain growth decreasing the average PAH size and the overall abundance of carbonaceous grains in low metallicity environments. The physical conditions, especially the high temperatures, in the central apertures of ESO137-G034 are likely to result in the inhibited grain growth as in low metallicity SF systems, leading to a high fraction of neutral PAHs with very small sizes.

However, the central regions of all our targets tend to have supersolar rather than sub-solar metallicity, according to a rough estimate of the metallicity based on the commonly used metallicity-sensitive line ratio $\rm \frac{[O~{\footnotesize III}]/H\beta}{[N~{\footnotesize II}]/H\alpha}$ measured by \cite{Davies.etal.2020} and the metallicity diagnostic calibration by \cite{Kewley.etal.2019}. Therefore, if the inhibited grain growth in the central region of ESO137-G034 is true, we need other explanations, rather the low metallicity, to account for the physical conditions in the central region of ESO137-G034. To ascertain the potential role played by star formation in affecting the PAH properties in such AGN dominated systems, dedicated SED modeling is required, which deserves specific study. 

As detailed in Section~\ref{sec4}, we also find evidence of extra heating in ESO137-G034, which could be due to shocks of low velocities since we do not find significant evidence of fast radiative shocks therein (see Figure~\ref{Lineratio}). Accordingly, given the size distribution of astronomical PAHs has a narrow peak around $N_{\rm C} = 150$ in normal galaxies (\citealt{Weingartner&Draine2001, Draine&Li2007}) and PAHs will be only partially destroyed in shocks with $v < 100\,\rm km\,s^{-1}$ (\citealt{Micelotta.etal.2010}), an alternative explanation for the PAH characteristics of ESO137-G034 is that the PAH size distribution therein is changed via preferential erosion rather than complete destruction. This explanation is plausible as preferential erosion can result in a shift of the narrow peak in PAH size distribution to a location more efficiently radiating short-wavelength PAH features (i.e., shift toward smaller sizes), with an overall reduction in the amount of PAHs. 

This explanation also applies to the PAH characteristics of aperture 5~\&~6 in NGC\,3081 and consistent with their overall higher PAH 11.3\,$\mum$/7.7\,$\mum$ ratios as partially destroyed PAHs with irregular structures could exhibit higher PAH 11.3\,$\mum$/7.7\,$\mum$ ratios (e.g., \citealt{Li2020}), though this conclusion is still on debate (e.g., \citealt{Rigopoulou.etal.2024}). Figure~\ref{PAHratio} shows that some literature Seyferts have PAH properties similar to those of ESO-137-G034 and aperture 5~\&~6 in NGC\,3081, and these Seyferts (i.e., PAH $6.2\,\mum/7.7\,\mum > 0.3$ and PAH $11.3\,\mum/7.7\,\mum > 0.6$) are mostly from \cite{Diamond-Stanic&Rieke2010}. According to \cite{Diamond-Stanic&Rieke2010}, we find that these Seyferts also exhibit evidence of extra heating (i.e., enhanced H$_2$ emission) and relatively weaker high ionization line (i.e., [O~{\footnotesize IV}]). Such emission line properties are similar to those of ESO137-G034, suggesting the aforementioned explanation for PAH characteristics of ESO137-G034 also apply to more general situations. Low-ionization nuclear emission-line region galaxies (LINERs), which tend to show enhanced H$_2$ emission with overall higher PAH 11.3\,$\mum$/7.7\,$\mum$ ratios (\citealt{Zhang.etal.2022}), are good targets to further ascertain the effects of collisional shocks that are associated the kinetic mode AGN feedback in affecting PAH properties of AGN systems such as ESO137-G034. 

The severe photo-ionization, no matter from the shock precursor or from the AGN, provide a good explanation for why the central aperture of MCG-05-23-016 contains more ionized PAHs of very large sizes. Even without the collisional destruction by shocks, where PAHs with $N_{\rm C} < 200$ can be completely destroyed by shocks with velocity $\sim 150\,\rm km\,s^{-1}$ (\citealt{Micelotta.etal.2010}), the severe photo-ionization itself can contribute to the ionization of all PAHs and further photo-destruction of small PAHs (e.g., \citealt{Allain.etal.1996, Holm.etal.2011}). Therefore, the extra destruction effects associated with the (fast radiative) shock precursor in addition to that of AGN will naturally result in a higher fraction of ionized PAHs with large sizes. Moreover, the weak shielding of PAHs by hydrogen around the AGN of MCG-05-23-016 can further contribute to the modification of PAH properties, resulting in enhanced photo-ionization and then photo-destruction of PAHs.

The central kpc regions of more than 30\% of the literature Seyferts in Figure~\ref{PAHratio} were studied by \cite{Garcia-Bernete.etal.2022a} at a spatial resolution of $\sim 6\arcsec$. Specifically, we find based on the measurements by \cite{Garcia-Bernete.etal.2022a} that for those Seyfert nuclei exhibiting relatively weak PAH 6.2\,$\mum$ feature (i.e., PAH $6.2\,\mum/7.7\,\mum \lesssim 0.2$), their ionized emission line ratios are concentrated around the distribution peak of the shock model results (i.e., the green contours in Figure~\ref{Lineratio}).\footnote{PAH measurements of upper limits are excluded from the analysis and the same for emission line measurements of upper limits.} This is not the case for those Seyfert nuclei  of relatively strong PAH 6.2\,$\mum$ feature. These results support the scenario that the more ionized PAHs of large sizes in AGN systems such as MCG-05-23-016 is due to severe photo-ionization from the (fast radiative) shock precursor in addition to the AGN. Moreover, as shown in Figure~\ref{PAHratio}, some low-redshift quasars also exhibit very low ratios of PAH $6.2\,\mum/7.7\,\mum$ (i.e., $\lesssim 0.15$; \citealt{Xie&Ho2022}). These quasars constitute a good sample to investigate whether the results obtained here for a single case apply to more extreme AGN systems, which are dominated by the radiative mode AGN feedback. 

Aperture 2 of NGC\,3081 contains more ionized PAHs of large sizes as well (see Figure~\ref{PAHratio}), while this aperture does not exhibit similar emission line ratios as the central aperture of MCG-05-23-016 (see Figure~\ref{Lineratio}). This is plausibly due to the dilution effect of nuclear line emission, as we find that some peripheral apertures of smaller sizes overlapping with aperture 2 of NGC\,3081 exhibit similar emission line ratios as the central aperture of MCG-05-23-016. Aperture 2 of NGC\,3081 also covers the end of the nuclear radio jet found in this object, an indirect evidence for fast shocks (see \citealt{Zhang.etal.2024}). Furthermore, in some testing work not shown here, we find that apertures of smaller sizes in NGC\,3081 containing more ionized PAHs of large sizes appear to lie behind regions showing emission line characteristics of fast radiative shocks. This tentative result suggests that the destruction of small PAHs via fast radiative shocks primarily occurs in the post-shock regions (see also \citealt{Donnan.etal.2023a}), and more importantly the modification of PAH properties is more like a long-duration process. This result worths specific study while beyond the scope here.

PAH features are supposed to be powerful diagnostics of different physical conditions, and should be better leveraged except for as indicators of star formation rates and molecular gas content. The analysis reported here is among the first attempts to quantitatively associate PAH characteristics with underlying physical conditions of different AGN environments. Based on our findings, it seems that, PAH characteristics, if well calibrated, have the potential to be powerful diagnostics of not only AGN activity but also different evolutionary processes associated with AGN. As part of a full set of diagnostics covering ionized and molecular gas, and more importantly tracing different evolutionary processes with distinct time scales, PAH features along with other emission lines can provide important quantitative information about the environments and physical conditions in which they are produced and processed.

Accordingly, combining the PAH characteristics and other emission line diagnostics covered by JWST spectra, we are able to have a more comprehensive understanding of different feedback effects of AGN activity, which is pivotal for fully understanding the galaxy formation and evolution. Further study, similar to the analysis done here, of a larger sample and for different AGN systems (e.g., quasars, LINERs, and Seyferts) could allow us to use PAH characteristics as diagnostics of different evolutionary processes in galaxies, which has a promising application in the era of JWST.

\section{Summary and Conclusions}\label{sec6}

Leveraging high quality JWST MIRI/MRS IFU observations, this letter showcases the distinct PAH characteristics in a sample of three Seyferts, based on the PAH measurements from a series of $3\arcsec\times3\arcsec$ (i.e., $\sim$ 500 pc) apertures (Section~\ref{sec2}). We find that positions occupied by apertures in the three targets on the PAH diagram are different from those occupied by regions in star-forming galaxies. Specifically, we find larger PAH 11.3\,$\mum$/7.7\,$\mum$ ratios or smaller PAH 6.2\,$\mum$/7.7\,$\mum$ ratios for apertures in the three targets, indicating overall more neutral or larger PAHs therein compared to regions in star-forming galaxies (Section~\ref{sec3}). In addition to the PAH diagram, we also present and discuss other emission line diagnostics, revealing the existence of extra heating or potential fast radiative shocks in these targets (Section~\ref{sec4}). Combining the PAH diagram and the emission line diagnostics, we further discuss the underlying mechanisms responsible for the PAH characteristics of the three targets (Section~\ref{sec5}).

The main findings of this work from the observations and model results can be summarized as follows:

\begin{enumerate}

\item The central regions exhibiting relatively strong PAH 6.2\,$\mum$ as well as 11.3\,$\mum$ features (i.e., PAH $6.2\,\mum/7.7\,\mum > 0.3$ and PAH $11.3\,\mum/7.7\,\mum > 0.4$) contain a high fraction of neutral PAHs with small sizes. Such PAH characteristics as in ESO137-G034 reflect a shift of PAH size distribution toward small sizes with an overall reduced amount of PAHs, which can be explained by inhibited growth or preferential erosion of PAHs under the specific environments. The latter can be attributed to the collisional shock heating associated with the kinetic mode AGN feedback.

\item  The central regions exhibiting relatively weak PAH 6.2\,$\mum$ feature (i.e., PAH $6.2\,\mum/7.7\,\mum \lesssim 0.2$) contain a high fraction of ionized PAHs with large sizes. Such PAH characteristics as in the central region of MCG-05-23-016 are plausibly due to the severe photo-ionization from the (fast radiative) shock precursor in addition to the AGN under a environment of weak shielding of PAHs, which can contribute to the ionization of all PAHs and further destruction of small PAHs. These results imply that the effects associated with radiative mode AGN feedback are regulating the PAH properties in such environments.

\item The central regions of most Seyferts, including NGC\,3081 studied here, overall have a wide distribution of PAH band ratios, covering the values as in ESO137-G034 to that of MCG-05-23-016. Specifically, Seyferts with relatively strong PAH 6.2\,$\mum$ as well as 11.3\,$\mum$ features and Seyferts with  relatively weak PAH 6.2\,$\mum$ feature have emission line properties similar to those of ESO137-G034 and MCG-05-23-016, respectively. This result suggests that the physical mechanisms used to explain the PAH properties of ESO137-G034 and MCG-05-23-016 may also apply generally.


\end{enumerate}

Based on our findings, it is promising to use PAH characteristics, after well calibrated based on further large sample analysis, in diagnosing different evolutionary processes associated with AGN activity. 

\newpage

\acknowledgements
{\footnotesize We thank the anonymous referee for helpful comments. This letter is part of a series from the Galaxy Activity, Torus, and Out- flow Survey (GATOS) collaboration. LZ, CP, EKSH, and MTL acknowledge grant support from the Space Telescope Science Institute (ID: JWST-GO-01670.007-A). IGB is supported by the Programa Atracci\'on de Talento Investigador ``C\'esar Nombela'' via grant 2023-T1/TEC-29030 funded by the Community of Madrid. IGB and DR acknowledges support from STFC through grants ST/S000488/1 and ST/W000903/1. AAH, and LHM acknowledges financial support by the grant PID2021-124665NB-I00 funded by the Spanish Ministry of Science and Innovation and the State Agency of Research MCIN/AEI/10.13039/501100011033 PID2021-124665NB-I00 and ERDF A way of making Europe. CRA, AA and DEA acknowledge support by the project PID2022-141105NB-I00 ``Tracking active galactic nuclei feedback from parsec to kilo-parsec scales'', funded by MICINN-AEI/10.13039/501100011033. MPS acknowledges support from grants RYC2021-033094-I and CNS2023-145506 funded by MCIN/AEI/10.13039/501100011033 and the European Union NextGenerationEU/PRTR. CR acknowledges support from Fondecyt Regular grant 1230345 and ANID BASAL project FB210003. AJB acknowledges funding from the “FirstGalaxies” Advanced Grant from the European Research Council (ERC) under the European Union’s Horizon 2020 research and innovation program (Grant agreement No. 789056). SGB acknowledges support from the Spanish grant PID2022-138560NB-I00, funded by MCIN/AEI/10.13039/501100011033/FEDER, EU.  MS acknowledges support by the Ministry of Science, Technological Development and Innovation of the Republic of Serbia (MSTDIRS) through contract no. 451-03-66/2024-03/200002 with the Astronomical Observatory (Belgrade). EB acknowledges support from the Spanish grant PID2022-138621NB-I00, funded by MCIN/AEI/10.13039/501100011033/FEDER, EU. EL-R is supported by the NASA/DLR Stratospheric Observatory for Infrared Astronomy (SOFIA) under the 08\_0012 Program.  SOFIA is jointly operated by the Universities Space Research Association,Inc.(USRA), under NASA contract NNA17BF53C, and the Deutsches SOFIA Institut (DSI) under DLR contract 50OK0901 to the University of Stuttgart. EL-R is also supported by the NASA Astrophysics Decadal Survey Precursor Science (ADSPS) Program (NNH22ZDA001N-ADSPS) with ID 22-ADSPS22-0009 and agreement number 80NSSC23K1585. MJW acknowledges support from a Leverhulme Emeritus Fellowship, EM-2021-064. This work is based on observations made with the NASA/ESA/CSA James Webb Space Telescope. The data were obtained from the Mikulski Archive for Space Telescopes at the Space Telescope Science Institute, which is operated by the Association of Universities for Research in Astronomy, Inc., under NASA contract NAS 5-03127 for JWST. The specific observations analyzed here can be accessed via \dataset[doi: 10.17909/vre3-m991]{https://doi.org/10.17909/vre3-m991}.}


\appendix

\section{Measurements and Figures}\label{secA}
Tables for measurements of PAH features and emission lines involved in this work, as well as ancillary figures discussed in the main text.

\startlongtable
\tablenum{A1}
\setlength{\tabcolsep}{10pt}
\begin{deluxetable*}{cccc}
\tablecolumns{4}
\tablecaption{Measurements of PAH Features}
\tablehead{
\colhead{Region} & \colhead{log $f_{\rm PAH}^{6.2}$} & \colhead{log $f_{\rm PAH}^{7.7}$} & \colhead{log $f_{\rm PAH}^{11.3}$}\\
\colhead{(-)} & \colhead{[$\rm erg/s/cm^{2}$]} & \colhead{[$\rm erg/s/cm^{2}$]} & \colhead{[$\rm erg/s/cm^{2}$]} \\
\colhead{(1)} & \colhead{(2)} & \colhead{(3)} & \colhead{(4)}}
\startdata
ESO137-G034\_reg1    & $-$13.05  $\pm$   0.13    & $-$12.61  $\pm$   0.11     & $-$12.82   $\pm$    0.08 \\
ESO137-G034\_reg2    & $-$13.33  $\pm$   0.06    & $-$13.03  $\pm$   0.09     & $-$13.12   $\pm$    0.05 \\
ESO137-G034\_reg3    & $-$13.24  $\pm$   0.05    & $-$12.75  $\pm$   0.05     & $-$13.10   $\pm$    0.05 \\
MCG-05-23-016\_reg1    & $-$13.18  $\pm$   0.37    & $-$11.96  $\pm$   0.18     & $-$12.39   $\pm$    0.24 \\
NGC3081\_reg1    & $-$13.16  $\pm$   0.23    & $-$12.57  $\pm$   0.18     & $-$12.91   $\pm$    0.17 \\
NGC3081\_reg2    & $-$13.40  $\pm$   0.09    & $-$12.52  $\pm$   0.04     & $-$13.07   $\pm$    0.07 \\
NGC3081\_reg3    & $-$13.12  $\pm$   0.07    & $-$12.50  $\pm$   0.06     & $-$13.08   $\pm$    0.06 \\
NGC3081\_reg4    & $-$13.18  $\pm$   0.03    & $-$12.47  $\pm$   0.02     & $-$12.94   $\pm$    0.02 \\
NGC3081\_reg5    & $-$13.40  $\pm$   0.04    & $-$12.94  $\pm$   0.04     & $-$13.13   $\pm$    0.03 \\
NGC3081\_reg6    & $-$13.35  $\pm$   0.06    & $-$12.98  $\pm$   0.09     & $-$13.13   $\pm$    0.08 \\
NGC3081\_reg7    & $-$13.32  $\pm$   0.02    & $-$12.78  $\pm$   0.03     & $-$13.14   $\pm$    0.04 \\
NGC3081\_reg8    & $-$13.28  $\pm$   0.03    & $-$12.69  $\pm$   0.03     & $-$12.99   $\pm$    0.03 \\
NGC3081\_reg9    & $-$13.38  $\pm$   0.03    & $-$12.84  $\pm$   0.04     & $-$13.09   $\pm$    0.03 \\
\enddata
\tablecomments{\footnotesize Flux of PAH features.}
\label{tabPAHs}
\end{deluxetable*}

\startlongtable
\tablenum{A2}
\setlength{\tabcolsep}{3pt}
\begin{deluxetable*}{ccccccc}
\tablecolumns{7}
\tablecaption{Measurements of Ionized Emission Lines}
\tablehead{
\colhead{Region} & \colhead{log $f_{\rm [Ne~V]}$} & \colhead{log $f_{\rm [Ne~III]}$} & \colhead{log $f_{\rm [Ne~II]}$} & \colhead{log $f_{\rm [O~IV]}$} & \colhead{log $f_{\rm [S~IV]}$} & \colhead{log $f_{\rm [Fe~II]}$} \\
\colhead{(-)} & \colhead{[$\rm erg/s/cm^{2}$]} & \colhead{[$\rm erg/s/cm^{2}$]} & \colhead{[$\rm erg/s/cm^{2}$]} & \colhead{[$\rm erg/s/cm^{2}$]} & \colhead{[$\rm erg/s/cm^{2}$]}  & \colhead{[$\rm erg/s/cm^{2}$]} \\
\colhead{(1)} & \colhead{(2)} & \colhead{(3)} & \colhead{(4)} & \colhead{(5)} & \colhead{(6)} & \colhead{(7)}}
\startdata
ESO137-G034\_reg1     & $-$12.72    $\pm$    0.01       &   $-$12.32      $\pm$      0.01      &  $-$12.75     $\pm$     0.01     & $-$12.18    $\pm$    0.01     & $-$12.37    $\pm$    0.01      &  $-$13.31     $\pm$     0.02 \\
ESO137-G034\_reg2     & $-$12.96    $\pm$    0.01       &   $-$12.48      $\pm$      0.01      &  $-$12.91     $\pm$     0.01     & $-$12.39    $\pm$    0.01     & $-$12.58    $\pm$    0.01      &  $-$13.34     $\pm$     0.02 \\
ESO137-G034\_reg3     & $-$13.12    $\pm$    0.01       &   $-$12.61      $\pm$      0.01      &  $-$12.98     $\pm$     0.01     & $-$12.58    $\pm$    0.01     & $-$12.73    $\pm$    0.01      &  $-$13.36     $\pm$     0.03 \\
MCG-05-23-016\_reg1     & $-$12.91    $\pm$    0.01       &   $-$12.77      $\pm$      0.01      &  $-$12.75     $\pm$     0.01     & $-$12.72    $\pm$    0.02     & $-$13.13    $\pm$    0.03      &  $-$14.29     $\pm$     0.05 \\
NGC3081\_reg1     & $-$12.56    $\pm$    0.01       &   $-$12.48      $\pm$      0.01      &  $-$13.04     $\pm$     0.01     & $-$12.19    $\pm$    0.01     & $-$12.54    $\pm$    0.01      &  $-$14.28     $\pm$     0.03 \\
NGC3081\_reg2     & $-$13.12    $\pm$    0.01       &   $-$13.02      $\pm$      0.01      &  $-$13.64     $\pm$     0.01     & $-$12.69    $\pm$    0.01     & $-$13.16    $\pm$    0.01      &  $-$14.77     $\pm$     0.06 \\
NGC3081\_reg3     & $-$13.29    $\pm$    0.01       &   $-$13.29      $\pm$      0.01      &  $-$13.72     $\pm$     0.01     & $-$12.95    $\pm$    0.01     & $-$13.38    $\pm$    0.01      &  $-$14.74     $\pm$     0.03 \\
NGC3081\_reg4     & $-$13.27    $\pm$    0.02       &   $-$13.24      $\pm$      0.01      &  $-$13.71     $\pm$     0.01     & $-$12.91    $\pm$    0.02     & $-$13.37    $\pm$    0.01      &  $-$14.72     $\pm$     0.10 \\
NGC3081\_reg5     & $-$13.77    $\pm$    0.01       &   $-$13.57      $\pm$      0.01      &  $-$14.09     $\pm$     0.02     & $-$13.39    $\pm$    0.04     & $-$13.78    $\pm$    0.01      &  $-$14.84     $\pm$     0.26 \\
NGC3081\_reg6     & $-$13.07    $\pm$    0.01       &   $-$13.01      $\pm$      0.01      &  $-$13.63     $\pm$     0.01     & $-$12.71    $\pm$    0.01     & $-$13.03    $\pm$    0.01      &  $-$14.50     $\pm$     0.09 \\
NGC3081\_reg7     & $-$13.23    $\pm$    0.01       &   $-$13.26      $\pm$      0.01      &  $-$13.90     $\pm$     0.02     & $-$12.89    $\pm$    0.01     & $-$13.23    $\pm$    0.01      &  $-$14.84     $\pm$     0.15 \\
NGC3081\_reg8     & $-$13.24    $\pm$    0.01       &   $-$13.24      $\pm$      0.01      &  $-$13.85     $\pm$     0.02     & $-$12.85    $\pm$    0.01     & $-$13.22    $\pm$    0.01      &  $-$14.85     $\pm$     0.14 \\
NGC3081\_reg9     & $-$13.77    $\pm$    0.02       &   $-$13.56      $\pm$      0.01      &  $-$14.08     $\pm$     0.01     & $-$13.29    $\pm$    0.01     & $-$13.70    $\pm$    0.01      &  $-$14.97     $\pm$     0.04 \\
\enddata
\tablecomments{\footnotesize Flux of ionized emission lines.}
\label{tabLines}
\end{deluxetable*}

\startlongtable
\tablenum{A3}
\setlength{\tabcolsep}{2pt}
\begin{deluxetable*}{ccccccccc}
\tabletypesize{\footnotesize}
\tablecolumns{9}
\tablecaption{Measurements of Hydrogen Rotational Emission Lines}
\tablehead{
\colhead{Region} & \colhead{log $f_{{\rm H}_{2}\,S(1)}$} & \colhead{log $f_{{\rm H}_{2}\,S(2)}$} & \colhead{log $f_{{\rm H}_{2}\,S(3)}$} & \colhead{log $f_{{\rm H}_{2}\,S(4)}$} & \colhead{log $f_{{\rm H}_{2}\,S(5)}$} & \colhead{log $f_{{\rm H}_{2}\,S(6)}$} & \colhead{log $f_{\rm Pf\alpha}$} & \colhead{log $f_{\rm Hu\alpha}$} \\
\colhead{(-)} & \colhead{[$\rm erg/s/cm^{2}$]} & \colhead{[$\rm erg/s/cm^{2}$]} & \colhead{[$\rm erg/s/cm^{2}$]} & \colhead{[$\rm erg/s/cm^{2}$]} & \colhead{[$\rm erg/s/cm^{2}$]} & \colhead{[$\rm erg/s/cm^{2}$]} & \colhead{[$\rm erg/s/cm^{2}$]} & \colhead{[$\rm erg/s/cm^{2}$]} \\
\colhead{(1)} & \colhead{(2)} & \colhead{(3)} & \colhead{(4)} & \colhead{(5)} & \colhead{(6)} & \colhead{(7)} & \colhead{(8)} & \colhead{(9)} }
\startdata
ESO137-G034\_reg1       &  $-$13.77     $\pm$     0.07       &  $-$13.76     $\pm$     0.02       &  $-$13.21     $\pm$     0.01       &  $-$13.72     $\pm$     0.01       &  $-$13.35     $\pm$     0.01       &  $-$14.00     $\pm$     0.01          &     $-$14.15        $\pm$        0.03          &     $-$14.55        $\pm$        0.04 \\
ESO137-G034\_reg2       &  $-$14.01     $\pm$     0.03       &  $-$14.05     $\pm$     0.02       &  $-$13.49     $\pm$     0.01       &  $-$13.99     $\pm$     0.01       &  $-$13.61     $\pm$     0.01       &  $-$14.20     $\pm$     0.01          &     $-$14.35        $\pm$        0.03          &     $-$14.74        $\pm$        0.04 \\
ESO137-G034\_reg3       &  $-$13.95     $\pm$     0.02       &  $-$14.00     $\pm$     0.01       &  $-$13.44     $\pm$     0.01       &  $-$13.95     $\pm$     0.01       &  $-$13.56     $\pm$     0.01       &  $-$14.20     $\pm$     0.01          &     $-$14.43        $\pm$        0.04          &     $-$14.93        $\pm$        0.04 \\
MCG-05-23-016\_reg1       &  $-$13.85     $\pm$     0.06       &  $-$14.15     $\pm$     0.05       &  $-$13.70     $\pm$     0.01       &  $-$14.23     $\pm$     0.03       &  $-$13.92     $\pm$     0.03       &  $-$14.50     $\pm$     0.29          &     $-$14.02        $\pm$        0.08          &     $-$14.50        $\pm$        0.18 \\
NGC3081\_reg1       &  $-$13.63     $\pm$     0.01       &  $-$13.91     $\pm$     0.02       &  $-$13.54     $\pm$     0.01       &  $-$14.00     $\pm$     0.01       &  $-$13.70     $\pm$     0.01       &  $-$14.37     $\pm$     0.02          &     $-$14.40        $\pm$        0.03          &     $-$14.88        $\pm$        0.18 \\
NGC3081\_reg2       &  $-$14.00     $\pm$     0.01       &  $-$14.33     $\pm$     0.01       &  $-$14.06     $\pm$     0.01       &  $-$14.54     $\pm$     0.01       &  $-$14.26     $\pm$     0.01       &  $-$15.00     $\pm$     0.04          &     $-$15.06        $\pm$        0.07          &     $-$15.48        $\pm$        1.84 \\
NGC3081\_reg3       &  $-$14.13     $\pm$     0.01       &  $-$14.47     $\pm$     0.01       &  $-$14.32     $\pm$     0.01       &  $-$14.72     $\pm$     0.02       &  $-$14.39     $\pm$     0.02       &  $-$15.17     $\pm$     0.06          &     $-$15.03        $\pm$        0.05          &     $-$15.54        $\pm$        0.06 \\
NGC3081\_reg4       &  $-$13.90     $\pm$     0.01       &  $-$14.25     $\pm$     0.01       &  $-$14.06     $\pm$     0.01       &  $-$14.53     $\pm$     0.01       &  $-$14.21     $\pm$     0.01       &  $-$14.96     $\pm$     0.03          &     $-$15.02        $\pm$        0.05          &     $-$15.54        $\pm$        0.04 \\
NGC3081\_reg5       &  $-$14.10     $\pm$     0.02       &  $-$14.51     $\pm$     0.02       &  $-$14.23     $\pm$     0.02       &  $-$14.78     $\pm$     0.03       &  $-$14.39     $\pm$     0.02       &  $-$15.09     $\pm$     0.05          &     $-$15.52        $\pm$        0.12          &     $-$16.10        $\pm$        0.13 \\
NGC3081\_reg6       &  $-$14.12     $\pm$     0.01       &  $-$14.43     $\pm$     0.01       &  $-$14.11     $\pm$     0.02       &  $-$14.64     $\pm$     0.02       &  $-$14.29     $\pm$     0.01       &  $-$14.98     $\pm$     0.07          &     $-$15.01        $\pm$        0.04          &     $-$15.38        $\pm$        0.07 \\
NGC3081\_reg7       &  $-$14.22     $\pm$     0.02       &  $-$14.60     $\pm$     0.02       &  $-$14.34     $\pm$     0.03       &  $-$14.88     $\pm$     0.03       &  $-$14.53     $\pm$     0.02       &  $-$15.29     $\pm$     0.09          &     $-$15.17        $\pm$        0.03          &     $-$15.56        $\pm$        0.09 \\
NGC3081\_reg8       &  $-$14.03     $\pm$     0.01       &  $-$14.38     $\pm$     0.01       &  $-$14.11     $\pm$     0.02       &  $-$14.65     $\pm$     0.02       &  $-$14.33     $\pm$     0.02       &  $-$15.12     $\pm$     0.05          &     $-$15.16        $\pm$        0.03          &     $-$15.65        $\pm$        0.28 \\
NGC3081\_reg9       &  $-$14.23     $\pm$     0.01       &  $-$14.58     $\pm$     0.01       &  $-$14.34     $\pm$     0.01       &  $-$14.85     $\pm$     0.03       &  $-$14.54     $\pm$     0.01       &  $-$15.37     $\pm$     0.08          &     $-$15.56        $\pm$        0.09          &     $-$15.72        $\pm$        1.25 \\
\enddata
\tablecomments{\footnotesize Flux of hydrogen emission lines.}
\label{tabH2s}
\end{deluxetable*}

\startlongtable
\tablenum{A4}
\setlength{\tabcolsep}{10pt}
\begin{deluxetable*}{ccccc}
\tablecolumns{5}
\tablecaption{Parameters Fitted from H$_{2}$ Rotational Lines}
\tablehead{
\colhead{Region} & \colhead{$n$} & \colhead{$T_{l}$} & \colhead{log $M_{\rm H_2}$} & \colhead{log $N_{\rm H_2}$}\\
\colhead{(-)} & \colhead{(-)} & \colhead{(K)} & \colhead{[$\rm M_{\odot}$]} & \colhead{[$\rm cm^{-2}$]} \\
\colhead{(1)} & \colhead{(2)} & \colhead{(3)} & \colhead{(4)} & \colhead{(5)}}
\startdata
ESO137-G034\_reg1 & 4.9 & 382 & 6.42 & 20.82 \\
ESO137-G034\_reg2 & 4.4 & 335 & 5.99 & 20.38 \\
ESO137-G034\_reg3 & 4.6 & 337 & 6.06 & 20.46 \\
MCG-05-23-016\_reg1 & 4.7 & 209 & 5.96 & 20.35 \\
NGC3081\_reg1 & 4.8 & 214 & 6.18 & 20.59 \\
NGC3081\_reg2 & 5.1 & 201 & 5.80 & 20.22 \\
NGC3081\_reg3 & 5.0 & 170 & 5.57 & 19.98 \\
NGC3081\_reg4 & 5.1 & 180 & 5.84 & 20.26 \\
NGC3081\_reg5 & 4.9 & 148 & 5.57 & 19.98 \\
NGC3081\_reg6 & 4.9 & 196 & 5.66 & 20.07 \\
NGC3081\_reg7 & 5.1 & 177 & 5.53 & 19.94 \\
NGC3081\_reg8 & 5.2 & 197 & 5.78 & 20.19 \\
NGC3081\_reg9 & 5.2 & 191 & 5.56 & 19.98 \\
\enddata
\tablecomments{\footnotesize Column (2) \& (3): The power-law index $n$ and low temperature end $T_{l}$ fitted from H$_{2}$ rotational lines. Column (4) \& (5): The derived mass and column density of H$_{2}$ at $T > 200\,\rm K$ for each aperture.}
\label{tabH2curves}
\end{deluxetable*}


\begin{figure*}[!ht]
\figurenum{A1}
\center{\includegraphics[width=0.85\linewidth]{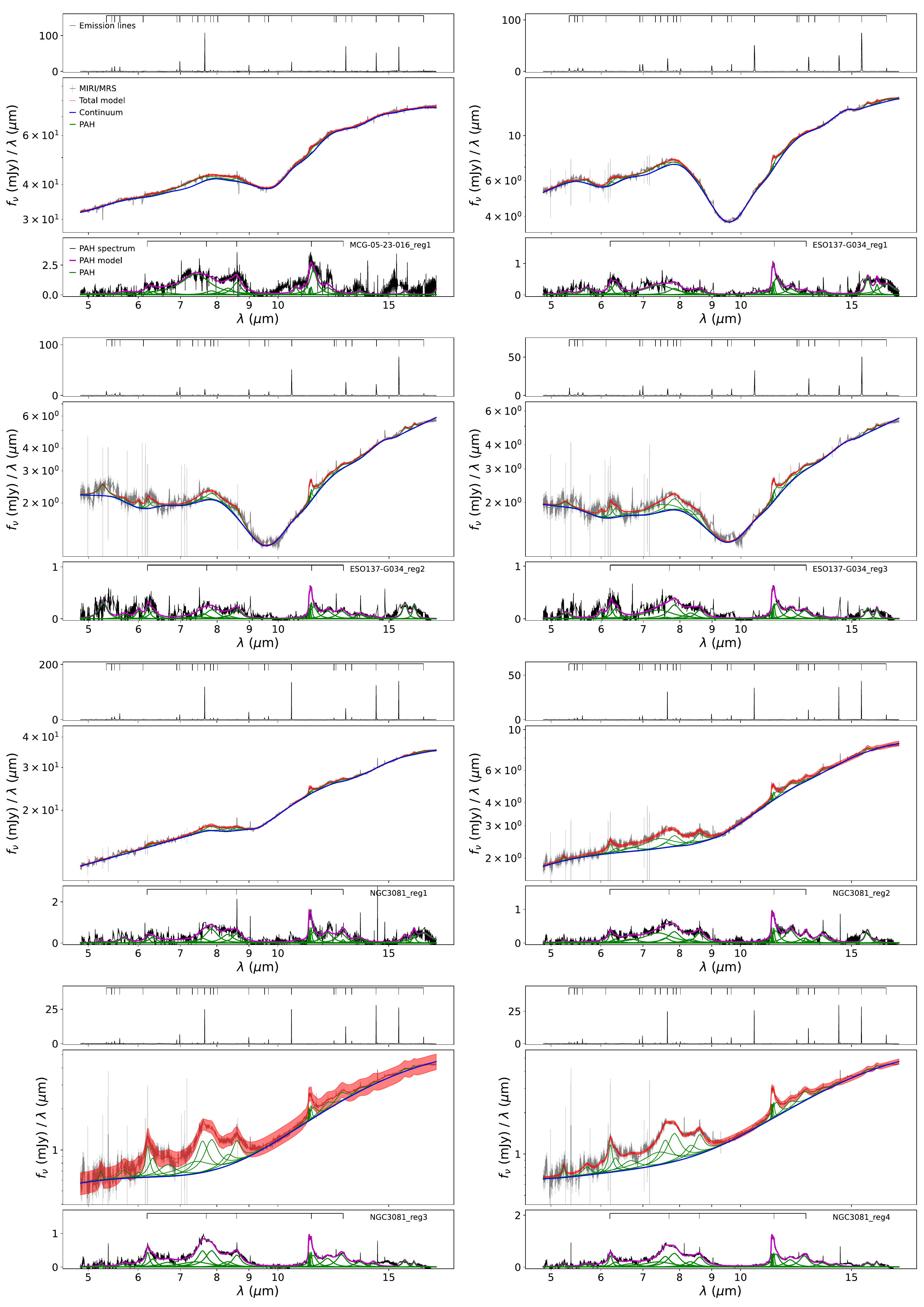}}
\end{figure*}

\begin{figure*}[!ht]
\figurenum{A1}
\center{\includegraphics[width=0.85\linewidth]{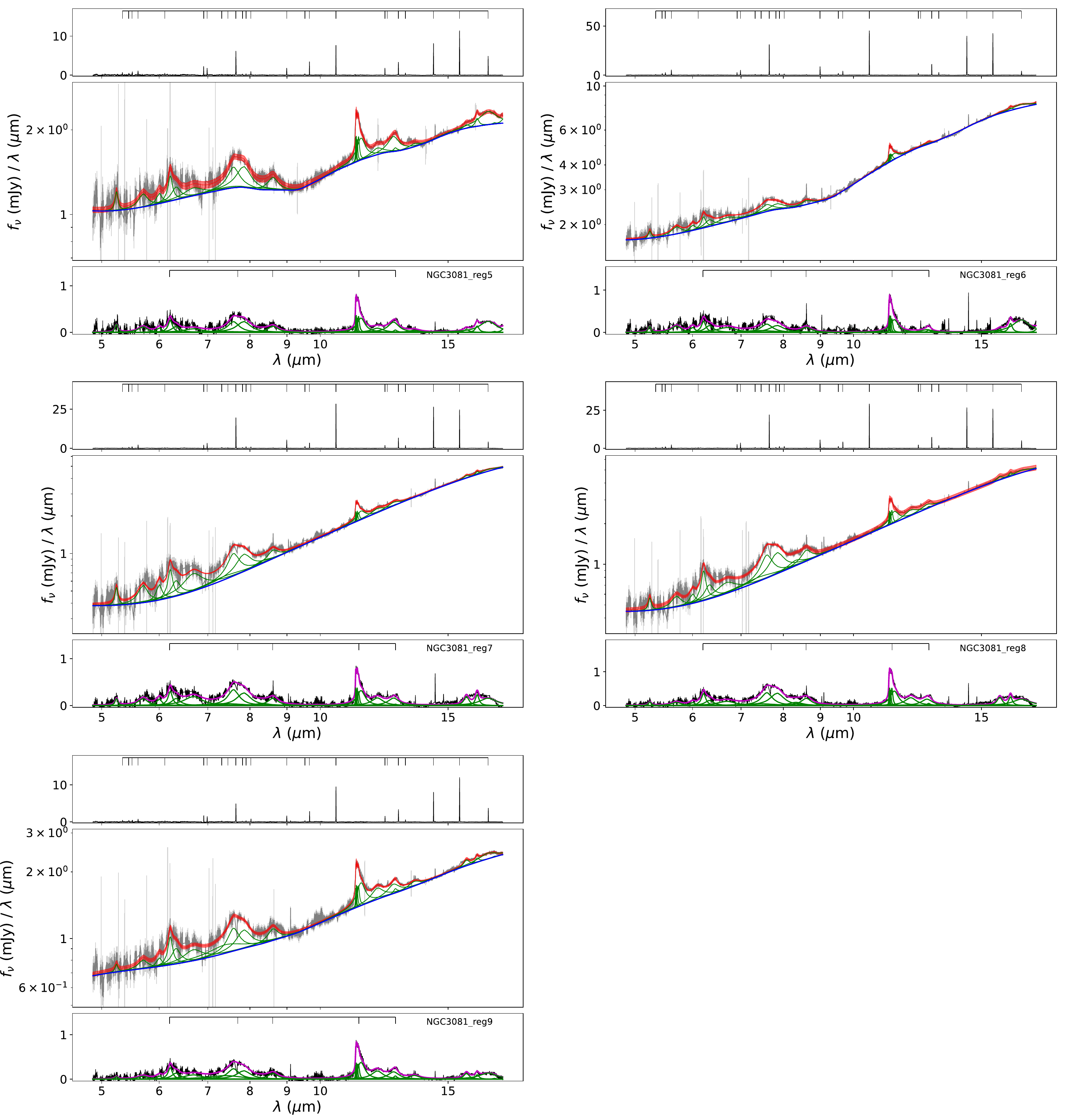}}
\caption{The same as Figure~\ref{Fit_Demo} but for all apertures studied here.}\label{Fit_Demos}
\end{figure*}

\begin{figure*}[!ht]
\figurenum{A2}
\center{\includegraphics[width=0.95\linewidth]{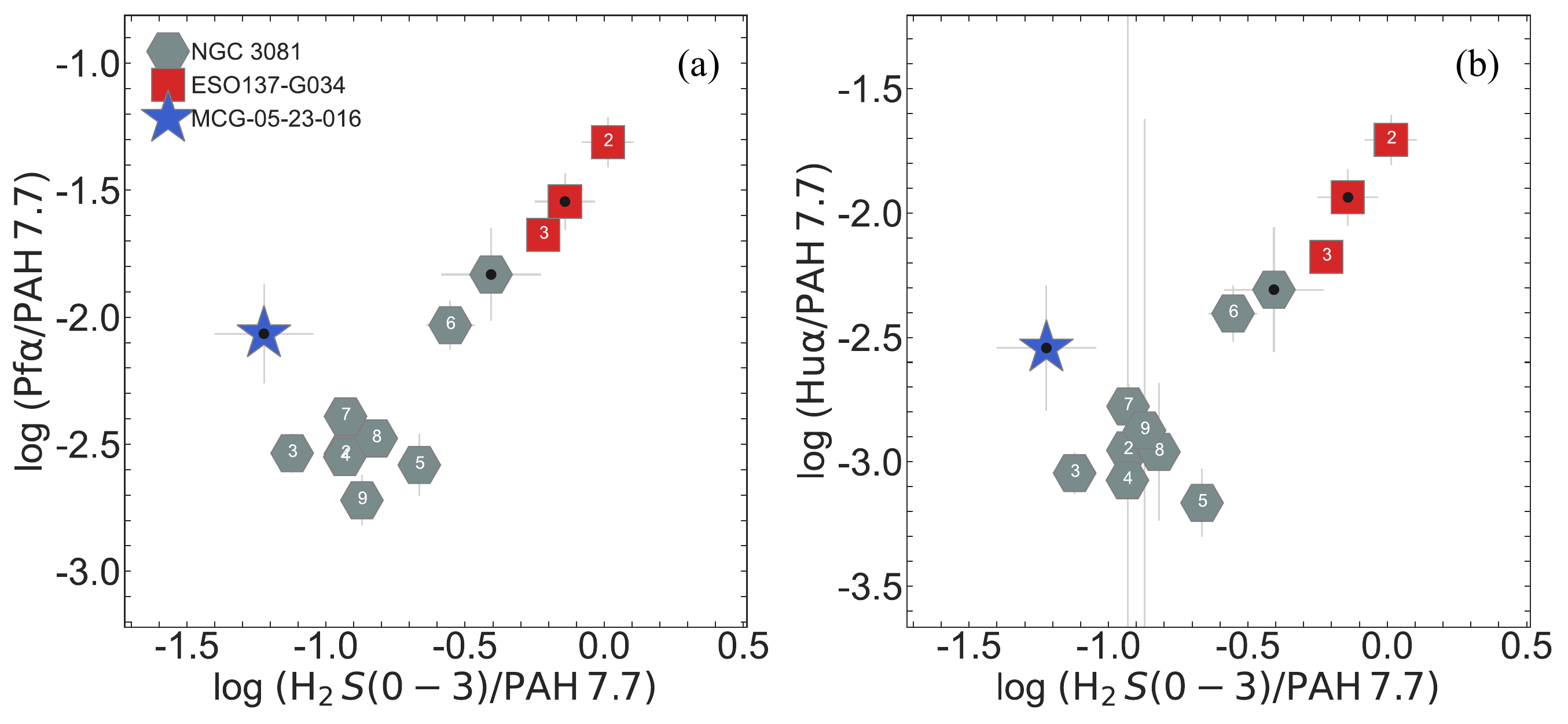}}
\caption{The correlation normalized by PAH 7.7\,$\mum$ emission between H$_{2}$ rotational emission and (a) atomic hydrogen Pf$\alpha$ emission and (b) atomic hydrogen Hu$\alpha$ emission, respectively, with markers the same as in Figure~\ref{PAHratio}.}\label{H_PAH}
\end{figure*}

\begin{figure*}[!ht]
\figurenum{A3}
\center{\includegraphics[width=0.95\linewidth]{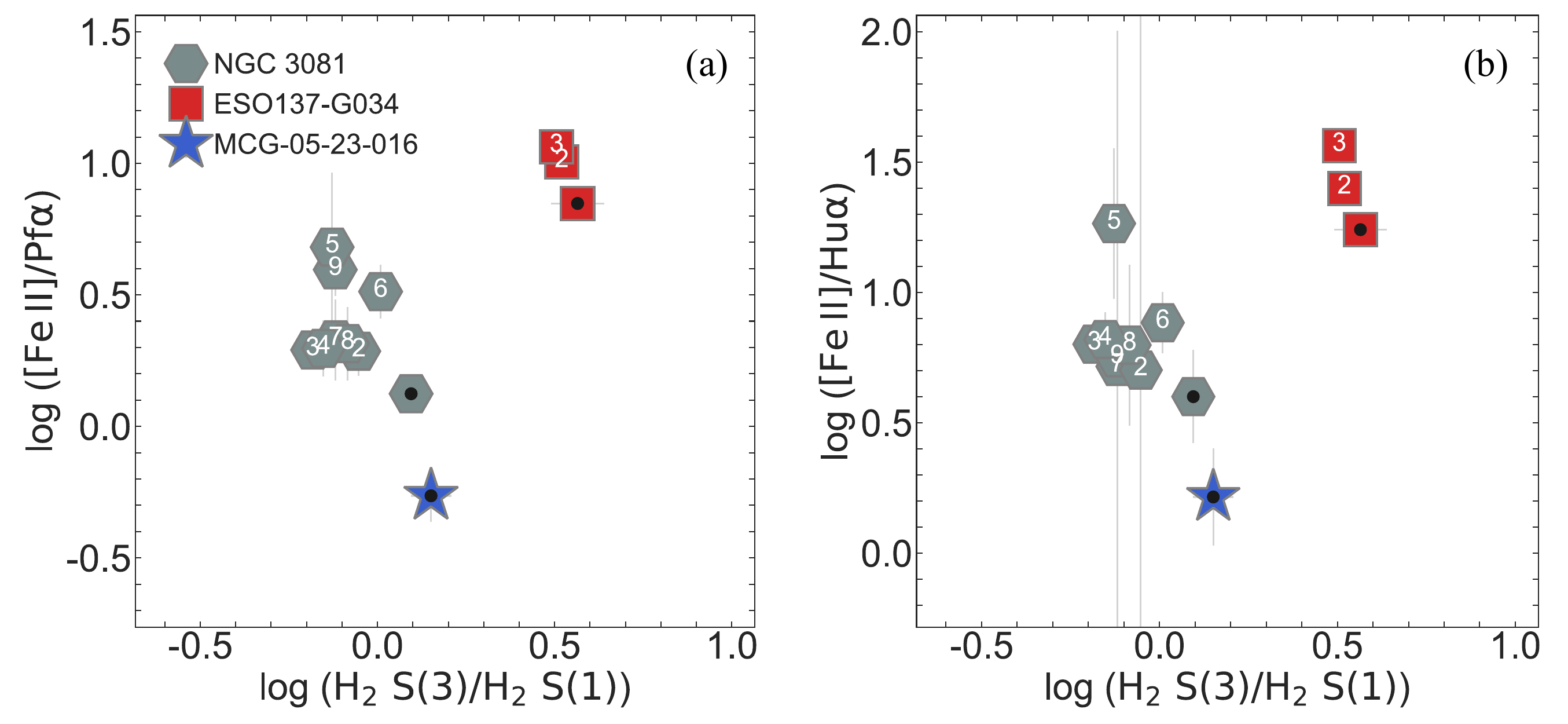}}
\caption{Diagnostic diagram of (a) [Fe~{\scriptsize II}]/Pf$\alpha$ and (b) [Fe~{\scriptsize II}]/Hu$\alpha$ versus ${\rm H_{2}}\,S(3)/{\rm H_{2}}\,S(1)$ for apertures in the three targets with markers the same as in Figure~\ref{PAHratio}.}\label{H_Fe}
\end{figure*}

\end{document}